\documentclass[aip,jcp,amsmath,amssymb,reprint,superscriptaddress]{revtex4-2}
\usepackage[dvipdfmx]{graphicx}
\usepackage{epsf}
\usepackage{bm}
\usepackage{physics}
\usepackage{lipsum}
\usepackage[normalem]{ulem}
\usepackage{txfonts}
\usepackage{color}

\graphicspath{{./fig/}{../fig/}}

\begin{document}
\title{Graph neural network-based structural classification of glass-forming liquids and its interpretation via self-attention mechanism
}

\author{Kohei Yoshikawa}
\affiliation{Division of Chemical Engineering, Department of Materials Engineering Science, Graduate School of Engineering Science, The University of Osaka, Toyonaka, Osaka 560-8531, Japan}

\author{Kentaro Yano}
\affiliation{Division of Chemical Engineering, Department of Materials Engineering Science, Graduate School of Engineering Science, The University of Osaka, Toyonaka, Osaka 560-8531, Japan}

\author{Shota Goto}
\affiliation{Division of Chemical Engineering, Department of Materials Engineering Science, Graduate School of Engineering Science, The University of Osaka, Toyonaka, Osaka 560-8531, Japan}

\author{Kang Kim}
\email{kk@cheng.es.osaka-u.ac.jp}
\affiliation{Division of Chemical Engineering, Department of Materials Engineering Science, Graduate School of Engineering Science, The University of Osaka, Toyonaka, Osaka 560-8531, Japan}

\author{Nobuyuki Matubayasi}
\email{nobuyuki@cheng.es.osaka-u.ac.jp}
\affiliation{Division of Chemical Engineering, Department of Materials Engineering Science, Graduate School of Engineering Science, The University of Osaka, Toyonaka, Osaka 560-8531, Japan}

\date{\today}

\begin{abstract}
Glass-forming liquids exhibit slow dynamics below their melting
 temperatures, maintaining an amorphous structure reminiscent of
 normal liquids. 
Distinguishing microscopic structures in the
 supercooled and high-temperature regimes remains a debated
 topic. 
Building on recent advances in machine learning, particularly
 Graph Neural Networks (GNNs), our study automatically extracts features,
 unveiling fundamental mechanisms
 driving structural changes at varying temperatures. 
We employ the self-attention mechanism to generate attention
 coefficients that quantify the
 importance of connections between graph nodes, providing insights into
 the rationale behind GNN predictions.
Exploring structural changes with decreasing temperature through the
 GNN+self-attention 
 using physically-defined structural descriptors, including the
 bond-orientational order parameter, Voronoi cell volume, and
 coordination number, we
 identify strong correlations between high attention coefficients and
 more disordered structures as a key indicator of 
 variations in glass-forming liquids. 
\end{abstract}
\maketitle

\section{Introduction}

When a liquid is cooled below its melting temperature, it enters a
supercooled state while retaining its disordered amorphous structure.
As the temperature continues to decrease toward the glass transition temperature,
the structural relaxation time experiences a rapid increase, eventually leading to 
solidification.
The phenomenon is commonly referred to as the glass
transition.
Theoretical and computational elucidation aimed unraveling the characteristics
that govern the slow dynamics in glass-forming liquids, particularly
within structures closely similar to those of normal liquids, has
presented a formidable problem.~\cite{berthier2011Theoretical}

The challenge of unraveling the structural order 
governing slow dynamics in glass-forming liquids 
has been met with recent breakthroughs.
In fact, this advancement employs
machine learning (ML) techniques, enhancing
predictive accuracy and unveiling fundamental mechanisms driving
structural changes.~\cite{liu2021Machine, jung2025Roadmap}
ML prove exceptional in detecting subtle
temperature-related structural changes, even with varying relaxation
times, revealing a crucial link between structural features and slow
dynamics in glass-forming liquids.

Specifically, ML methodologies have been applied for data obtained
from molecular dynamics (MD) simulations to
explore multifaceted origins of the slow dynamics from the atomistic
level.~\cite{ronhovde2011Detecting, cubuk2015Identifying,
schoenholz2016Structural,
cubuk2017Structureproperty, schoenholz2017Relationship,
sussman2017Disconnecting, sharp2018Machine, boattini2019Unsupervised,
bapst2020Unveiling, swanson2020Deep, richard2020Predicting,
landes2020Attractive, paret2020Assessing, 
boattini2020Autonomously, boattini2021Averaging, wang2021Inverse, zhang2022Machine, 
alkemade2022Comparing, coslovich2022Dimensionalitya, 
tripodo2022Neural, soltani2022Exploring, shiba2023BOTAN,
jung2023Predicting, alkemade2023Improvingb, 
oyama2023What, nguyen2023Automated, 
ciarella2023Dynamics, ciarella2023Finding, wu2023Unsupervised,
jiang2023Geometryenhanceda, kim2023Machine, pezzicoli2024Rotationequivariant,
jung2024Dynamic, jung2024Normalizing, janzen2024Classifying,
liu2024Classification, jiang2024Interpretinga, wang2024Predicting,
qiu2025Unsupervised, winter2025Glassy}
These investigations have effectively opened new perspectives for
comprehending the characteristics between liquids and glasses, spanning the
prediction of their properties to the elucidation of
structural and dynamical changes by varying temperatures.
The seamless integration of ML and MD 
represents a transformative approach, which
yields insights into the 
behaviors inherent to various glass-forming liquids.

More recently, there has been a notable increase in the application of
sophisticated ML techniques, including Convolutional Neural Networks (CNNs)~\cite{simonyan2015Very}
and Graph Neural Networks (GNNs)~\cite{scarselli2009Graph, wu2021Comprehensive}, as evidenced in literatures.
In particular, the GNN is a neural network model designed to exploit the
capabilities of ML within graph structure, where the information is
represented in terms of nodes and edges.
It iteratively updates information associated with the nodes and
edges present in the input graph through message passing. 
Therefore, the GNN eliminates the requirement for introducing features
manually, allowing automatic discovery of relevant structural information.
Bapst \textit{et al.} achieved remarkable success in accurately
predicting the dynamic propensity in a three-dimensional
glass-forming liquids from its static structure, utilizing the power of GNNs.~\cite{bapst2020Unveiling}
The dynamic propensity refers to the identification of 
single-particle displacements exhibiting higher mobility than the
average.~\cite{widmer-cooper2004How, widmer-cooper2006Predicting, widmer-cooper2008Irreversible}
Consequently, their work established a compelling link between the dynamics
and
structural attributes of glass-forming liquids.
Shiba \textit{et al}. also succeeded in
predicting the dynamic propensity with higher accuracy than the results
of Bapst \textit{et al.}~\cite{shiba2023BOTAN}
Their success can be attributed to an extended approach that involved
learning not only individual particle displacements but also the
pair-particle displacements as part of the GNNs output.
Additionally, other advanced GNN techniques were described, including the
incorporation of a
geometry-enhanced GNN~\cite{jiang2023Geometryenhanceda} and a rotation-equivariant graph
structure.~\cite{pezzicoli2024Rotationequivariant}
These studies demonstrated that the GNNs 
can automatically extract features from 
graph-based inputs through
message passing, without the need for predefined physical descriptors.
However, 
their interpretability remains limited due to their black-box nature.

Swanson \textit{et al.} pursued an alternative approach and found that 
CNNs and GNNs possess the capability to 
discriminate between liquid and glass structures across the glass transition
temperature $T_\mathrm{g}$ of the two-dimensional glass-forming
liquids.~\cite{swanson2020Deep}
Notably, 
the self-attention mechanism, a method within eXplainable
Artificial Intelligence (XAI), provides
insights into how GNNs differentiate liquids and glasses.
Originally introduced in Graph Attention Networks, the self-attention
mechanism quantifies the importance of a given node in relation to its
edge-connected neighbors within the graph
structure, providing interpretability to black-box nature of GNN.~\cite{velickovic2017Graph, velickovic2018graph}
Other studies have also highlighted the effectiveness of the self-attention
mechanism in ML, revealing 
connections between structure features and dynamic
heterogeneities.~\cite{jiang2023Geometryenhanceda, jung2024Dynamic, wang2024Predicting}

Furthermore, Oyama \textit{et al.} demonstrated comparable liquid-glass classification
using CNNs.~\cite{oyama2023What, liu2024Classification}
They further clarified the discriminative features underlying CNNs' predictions by
employing the Gradient-weighted Class Activation Mapping (Grad-Cam)
method.
The Grad-Cam approach provides 
a local-explanation model for CNN
predictions,~\cite{selvaraju2020GradCAM} revealing 
a correlation between 
the spatial distribution of the Grad-Cam score
and that of the dynamic propensity.

\begin{figure*}[t]
\centering
\includegraphics[width=\textwidth]{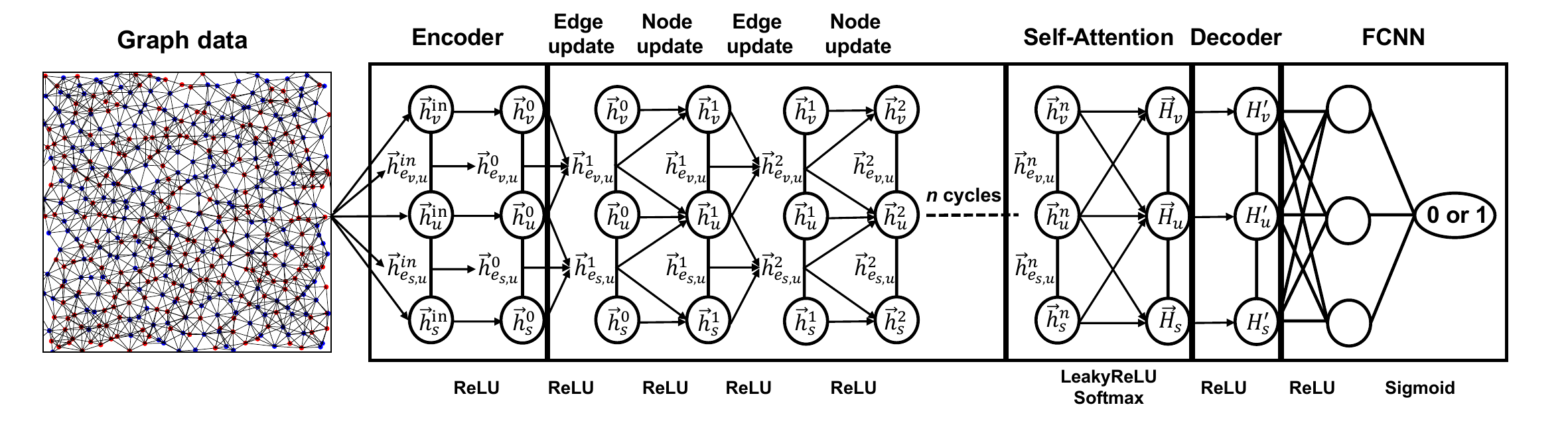}
\caption{Schematic picture of the GNN architecture conducted in this
 study.
The node and edge features, $\vec{h}^{\mathrm{in}}_v$ and
 $\vec{h}^{\mathrm{in}}_{e_{v,u}}$ are first generated from the graph
 data of the particle configuration obtained through MD simulations.
The GNN architecture includes the encoder, $n$ cycles of edge and node
 updates, self-attention layer, decoder, and FCNN for 
 predicting the binary classification between 0 (representing temperature $T_1$) and
 1 (representing temperature $T_2$).
The index $v$, $u$, $s$ refer to node indices within the range of 1 to
 $N$, where $N$ represents the total number of particles.
In this figure, 3 nodes are written, while in the calculations
 actually performed, the number of nodes is $N=4096$.}
\label{fig:GNN}
\end{figure*}

It is worth noting that 
the aforementioned studies have indeed
showcased the
success of ML and XAI in learning structural features in glass-forming
liquids.
However, GNNs have yet to fully capture the temperature-dependent
structural changes of glass-forming liquids, particularly for
three-dimensional systems, or provide a comprehensive
physical interpretation of the nodal features generated from GNNs.

In the present study, we employ
GNNs to classify the structures of 
a three-dimensional glass-forming liquid
model consisting of two types (type 1 and type 2) of particles of different
sizes.\cite{bernu1985Molecular, bernu1987Softsphere,
yamamoto1998Dynamics}
Our objective is to distinguish two temperature states
based solely on particle configurations, 
along the supercooled liquid
branch, which lies above the glass transition temperature.
In this regard, our approach differs from previous studies by Swanson
\textit{et al.} and Oyama \textit{et al.}, which used a constant
cooling rate distinguishing liquid and glass
structures.~\cite{swanson2020Deep, oyama2023What, liu2024Classification}

We further explore the rationale behind GNN predictions by
adopting the self-attention mechanism, with the aim of elucidating the
distinctive structural changes that occur as temperature varies.
Specifically, we compare node features generated by GNNs
with physically-defined structural descriptors, including bond
orientational orders (BOOs),~\cite{steinhardt1983Bondorientationala, tanaka2012Bond,
royall2015Role, tanaka2022Roles} Voronoi
cell volume,~\cite{voronoi1908Nouvelles} and coordination number (C.N.), to interpret
how graph-based ML differentiates the structures of
glass-forming liquids at different temperatures.
These structural descriptors
characterize the nearest neighbor particle environment, quantifying the
degree of the crystalline order.
These parameters are
expected to serve as salient
indicators governing the slow dynamics in glass-forming liquids.~\cite{tanaka2019Revealing}
Additionally, we compare the attention coefficients, which quantify the
importance of a target node relative to its neighboring nodes, with these
structural descriptors to provide a physical interpretation of the underlying
mechanism in GNNs predictions.



\section{Methods}

\subsection{Molecular dynamics simulations}
\label{method:MD}

The model employed was a well-known model of glass-forming liquids,
namely,
the three-dimensional binary soft-sphere (SS) model.~\cite{bernu1985Molecular,
bernu1987Softsphere, yamamoto1998Dynamics}
In the SS model, the interaction potential is based on
\begin{equation}
\phi(r) =
 \varepsilon_{\alpha\beta}\left(\frac{\sigma_{\alpha\beta}}{r}\right)^{12},
\end{equation}
and the truncated and force-shifted interaction, defined as:
\begin{equation}
U(r) = \phi(r) - \phi(r_\mathrm{c}) - (r-r_\mathrm{c}) \left.\frac{d\phi(r)}{dr} \right|_{r=r_\mathrm{c}},
\end{equation}
is employed to ensure 
that both the potential and the force approach zero continuously at the cutoff
length $r_\mathrm{c}=2.5\sigma_{\alpha\beta}$.
The size ratios are defined as $\sigma_{22}/\sigma_{11}= 1.2$ and
$\sigma_{12}/\sigma_{11} = 1.1$.
Additionally, the mass ratio is represented as $m_2/m_1=2$, while 
the equal energy scale is used with 
$\varepsilon_{11}=\varepsilon_{12}=\varepsilon_{22}$.
The reduced units for length, temperature, and time are 
$\sigma_{11}$, $\varepsilon_{11}/k_\mathrm{B}$,
and $\sqrt{m_1\sigma_{11}^2/\varepsilon_{11}}$, respectively.
The ratio of particle numbers between type 1 and type 2 is set to 50:50
within a total particle number of $N = 4096$.
The total number density is fixed at $\rho = 0.75$, and the
investigated temperatures are $T = 1.0$, $0.8$, $0.6$, $0.4$, $0.3$,
$0.25$, $0.24$, $0.23$, $0.22$, 
and $0.21$.
A time step $\Delta t=0.005$ was used.
All MD simulations were computed using Large-scale
Atomic/Molecular Massively Parallel Simulator (LAMMPS).~\cite{plimpton1995Fast}

The results of the radial distribution function $g(r)$ and
the self-part of the intermediate scattering function $F_s(k, t)$ are
found in Supplementary Fig.~S1(a) and S1(b) of the supplementary material,
respectively.
Furthermore, the temperature-dependent behavior of the structural
$\alpha$-relaxation time $\tau_\alpha$, which was determined from
$F_s(k, t)$, is illustrated in Fig.~S1(c) of the supplementary material.
The onset temperature $T_0$, below which $F_\mathrm{s}(k, t)$
begins to develop a two-step relaxation, was estimated to be $T_0 = 0.3$.~\cite{kim2013Multiple}
Furthermore, based on the power-law divergence behavior predicted 
by the mode-coupling theory, the critical temperature
was estimated as $T_\mathrm{c} = 0.198$.~\cite{kim2013Multiple}
Another glass transition temperature, identified as the divergence
temperature of $\tau_\alpha$, was estimated 
as $T_\mathrm{K}=0.181$ 
using the Vogel--Fulcher--Tammann equation,
 $\tau_\alpha = \tau_0 \exp(D/ (T-T_\mathrm{K}))$, as shown in
 Fig.~S1(c).
Thus, the temperature range examined in this study involves a deeply
supercooled regime above the glass transition temperature.
The temperature dependence of the potential energy is
displayed in Fig.~S1(d) of the supplementary material.

\subsection{GNN and self-attention mechanism}
\label{method:GNN}

The schematic picture of our GNN architecture is depicted in
Fig.~\ref{fig:GNN}.
The 
particle configuration can be represented by embedding features of
particle types and connections between pairs of particles
as attributes of nodes and edges of the graph structure, respectively. 
Specifically, 
we utilize the particle type (type 1 or type 2) as input feature vectors for the graph
nodes.
A pair of particles are treated as neighbors, forming
an edge when their distance is smaller than $r_\mathrm{cut}=1.8$.
The relative coordinates of the
pair of neighboring particles serve as input features for the edges.
The value of $r_\mathrm{cut}=1.8$ corresponds to 
the first minimum length of the radial distribution functions $g(r)$ of
large particles (type 2).
Note that this GNN approach is particularly appropriate when applied to many
particle systems since it effectively retains the three-dimensional structural information.

Consider a graph $G(V, E)$, where $V$ represents the set of nodes and
$E$ denotes the set of edges.
For any pair of nodes $v$ and $u$ belonging to $V$, an edge connecting
$v$ and $u$ is denoted as $e_{v, u}$.
The collection of nodes neighboring a node $v\in V$ is represented
as $N(v)$.
The initial step involves an encoding procedure using the input feature vectors
$\vec{h}^\mathrm{in}_v$ and
$\vec{h}^\mathrm{in}_{e_{v,u}}$ for a node $v\in V$ and an edge $e\in E$,
respectively.
This procedure is described as follows:
\begin{align}
\vec{h}^0_v &= \mathrm{EN}(\vec{h}^\mathrm{in}_v),\\
\vec{h}^0_{e_{v,u}} &= \mathrm{EN}(\vec{h}^\mathrm{in}_{e_{v,u}}),
\end{align}
where
$\mathrm{EN}(\cdot)$ represents the encoder function.
In practice, $\mathrm{EN}(\cdot)$ consists of 
one hidden layer and the output layer, both utilizing the Rectified Linear Unit
(ReLU)
activation function.
This encoder generates $\vec{h}^0_v$ and
$\vec{h}^0_{e_{v,u}}$, each composed of 
32 elements for node $v$ and edge $e_{v,u}$.

Secondly, 
features associated with nodes and edges within the graph are
mutually calculated, and updates to both node and edge features
undergo iteration for $n$ cycles.
This procedure is described as follows:
\begin{align}
\vec{h}^m_{e_{v,u}} &= \mathrm{MLP}\left(h^{m-1}_{e_{v,u}} \oplus
h^{m-1}_v \oplus h^{m-1}_u 
\right),\\
\vec{h}^m_v &= \mathrm{MLP}\left(\vec{h}^{m-1}_v \oplus
 \sum_{u\in N(v)} \vec{h}^m_{e_{v,u}}\right),
\end{align}
where $\mathrm{MLP}(\cdot)$ denote a multi-layer perception function,
consisting of one hidden layer and the output layer, both utilizing the
ReLU activation function.
Here, $m$ denotes the index of the cycle, \textit{i.e.}, $0 < m
\le n$, and 
the number of cycles was set to $n=10$.
This message passing procedure for updating nodes and edges propagates the
information from neighboring particles, eventually
encompassing the entire system.

We introduce the self-attention layer into our GNN architecture in order
to assess 
which node's
information to emphasize, considering the information exchanged among
neighboring nodes.~\cite{velickovic2017Graph, velickovic2018graph}
Note that Swanson \textit{et al.} employed the self-attention
layer at
the end of message passing in Ref.~\citenum{swanson2020Deep}, which is
compatible with our GNN architecture, as shown in Fig.~\ref{fig:GNN}.
The attention coefficient $e_{vu}$ is defined as
\begin{equation}
e_{vu} = a(\mathbf{W}\vec{h}^n_v, \mathbf{W}\vec{h}^n_u),
\end{equation}
which measures the importance of node $u$'s feature to node $v$.
Here, $\mathbf{W}$ is a learnable weight matrix and $a$ signifies the
so-called shared attentional mechanism.
In the procedure of $a$, the concatenation of two vectors,
$\mathbf{W}\vec{h}^n_v$ and $\mathbf{W}\vec{h}^n_u$, 
is fed into to a single-layer neural network, which is parameterized by a weight
vector $\vec{a}$, and further employs 
the Leaky ReLU nonlinearity (negative input slope of 0.2).
Using the Softmax function, the attention coefficient is normalized 
as follows:
\begin{equation}
\alpha_{vu}  = \mathrm{Softmax}(e_{vu}) = \frac{\exp(e_{vu})}{\sum_{s \in
\mathcal{N}(v)}\exp(e_{vs})}, 
\label{eq:attention}
\end{equation}
where $\mathcal{N}(v)$ represents the neighboring nodes of $v$, including node $v$ itself.
Note that $\alpha_{vu}$ 
is summed over $u$ to unity and 
is asymmetric with respect to the node
exchange, \textit{i.e.}, $\alpha_{vu}\ne \alpha_{uv}$ for $v \ne u$,
since the graph is directed.
Finally, new node features
\begin{equation}
\vec{H}_v  = \alpha_{vv} \mathbf{W} \vec{h}^n_v + \sum_{u\in \mathcal{N}(v)}
 \alpha_{vu} \mathbf{W} \vec{h}^n_u,
\end{equation}
are generated from the self-attention layer.

For each node, the output feature $\vec{H}_v$ with 32 elements is
transformed into a single element
variable $H'_v$ through the decoder that incorporates two hidden layers and 
employing the ReLU, and an output layer 
with the linear transformation.
The node feature $H'_v$ (4096 nodes) are combined and undergo training through
a fully connected neural network (FCNN).
This FCNN includes a single hidden layer 
utilizing the ReLU.

Through the encoder, message passing, self-attention layer, decoder, and
FCNN, a 
binary output predicts a value of 0 for the structure at $T_1$ and 1 at
$T_2$ following a sigmoid function.
In the context of the binary classification, the loss function employed
is binary cross-entropy, which is minimized during the training process
using the training dataset.

We used PyTorch for implementation, utilizing the Adam optimization algorithm.
The learning rate and batch size were set to $10^{-6}$ and 128,
respectively.
These hyperparameters were selected to ensure sufficient optimization of
the loss function, as discussed in the following paragraph.
The maximum number of epochs was set to 10000, and the training was
terminated when the loss function of the test dataset was
minimized, 
at which point the GNN was applied for prediction.

The number of weights in the GNN used in this study is 
4096 $\times$ 32
for node features and 
4096 $\times$ 32 $\times$
(number of coordination numbers per particle, ranging from 10 to 25) for edge features.
We obtained 500 equilibrium particle configurations at each
temperature from MD simulations.
Of these, 400 were allocated 
for training dataset, while the remaining 100 were used for
test dataset.
In total, 
each GNN model for a given $(T_1, T_2)$ combination utilized
800 configurations (400 from $T_1$ and 400 from $T_2$) used
for training, and 200 configurations (100 from $T_1$ and 100 from $T_2$)
for testing.

\subsection{Structural descriptors for local particle environment}
\label{method:FCNN}

The BOOs are useful to characterize the local
orientational orders of liquid and crystal
structures.~\cite{steinhardt1983Bondorientationala, tanaka2012Bond, royall2015Role, tanaka2022Roles}
To characterize the orientational symmetry among the neighbors of
particle $i$, the complex parameter is defined as
\begin{equation}
q_{lm} (i) = \frac{1}{N_\mathrm{b}(i)} \sum_{j=1}^{N_\mathrm{b}(i)}
 Y_{lm} (\hat{\bm{r}}_{ij}),
\end{equation}
where $N_\mathrm{b}(i)$ represents the number of first nearest neighboring particles of
particle $i$, within a cutoff radius of $r_\mathrm{cut}=1.8$.
This length corresponds to the length scale at which neighboring nodes
are connected by edges in the GNN.
Furthermore, 
$Y_{lm}(\cdot)$ are the spherical harmonics 
of degree $l$ and order $m$ (running from $m =-l$ to $m =+l$).
$\hat{\bm{r}}_{ij}$ represents the relative unit 
vector between particles $i$ and $j$.
To consider the rotational invariance, the local BOO is defined as follows:
\begin{equation}
Q_l(i) = \sqrt{\frac{4\pi}{2l +1}\sum_{m=-l}^l |q_{lm}(i)|^2}.
\end{equation}
While the selection of the order $l$ is arbitrary, $Q_6$
and $Q_4$ are often preferred due to its ability to differentiate between
liquid and crystal structures.
More specifically, $Q_6$ is sensitive to the 
hexagonal symmetry and tends to be large for 
FCC (face-centered
cubic) and HCP (hexagonal close-packed) structures, while $Q_4$ is
sensitive to structures with local cubic symmetry.
We employed the Pyboo code for the computation of $Q_6$ and $Q_4$.~\cite{Pyboo}

Note that the coarse-grained BOO has been proposed by averaging $q_{lm}(i)$ over
neighboring particles of particle $i$ and particle $i$ itself.~\cite{lechner2008Accurate}
This approach has been shown to provide a
more precise and discriminative characterization between
liquid and crystal structures when compared to the original BOO method.
We have evaluated the use of this coarse-grained BOO in our neural network;
however, we have opted not to employ it in this study.
Our observations indicated that the additional averaging step tends to
obscure the distinctions among disordered structural configurations
within the systems we are investigating.
Note that an analogous observation has been documented in a previous study, where unsupervised machine
learning was employed for dimension reduction of local structural
orders in glass-forming liquids.~\cite{coslovich2022Dimensionalitya}

An alternative approach to characterize local orders is based on Voronoi
tessellation.~\cite{voronoi1908Nouvelles}
In particular, the Voronoi cell serves as an indicator of
the local geometry
associated with each particle in glass-forming
liquids.~\cite{starr2002What, coslovich2007Understanding}
We conducted the Voronoi cell volume calculations using the Pyvoro code.~\cite{Pyvoro}
A final structural descriptor examined in this study is C.N. within the
first coordination shell, 
defined as the number of particles within $r_\mathrm{cut}=1.8$.

The temperature dependence of the probability density distributions for $Q_6$, $Q_4$,
Voronoi cell volume, and C.N. is shown in Fig.~S2(a), (b),
(c), and (d) of the supplementary material, respectively.
Figure~S2(a) and (b) reveal that $Q_6$
increases while $Q_4$ decreases as the
temperature decreases, indicating the development of hexagonal crystalline order.
Figure~S2(c) shows that 
the probability distribution of the Voronoi cell volume 
exhibits an approximately Gaussian shape at high temperatures.
However, 
as the temperature decreases, the distribution develops two distinct
peaks, indicating increasing spatial heterogeneity in the local structure.
Specifically, 
the distribution of small (type 1) and large (type 2) particles become distinct
as the supercooled liquid approaches the glass
transition.
Figure~S2(d) demonstrates 
C.N. tends to decrease slightly with decreasing the temperature.
Note that, although this result may seem counterintuitive, the value is
highly dependent on the choice of the cutoff distance, $r_\mathrm{cut}$.
A similar dependence on $r_\mathrm{cut}$ has been observed for BOO values.~\cite{mickel2013Shortcomings}

\begin{figure}[t]
\centering
\includegraphics[width=0.3\textwidth]{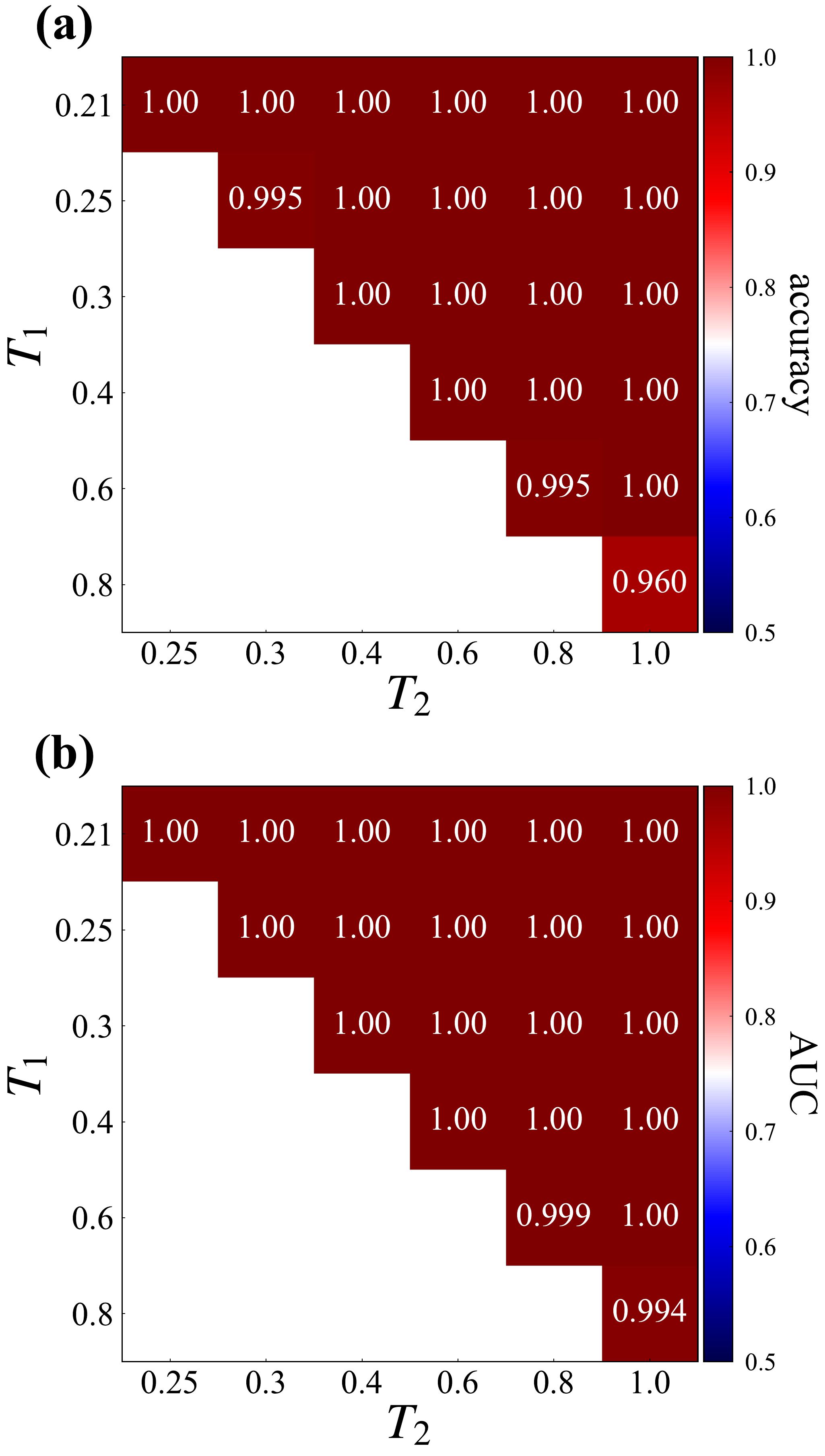}
\caption{Fraction of GNN
 prediction accuracy (a) and AUC (b) for classification between two
 temperatures, $T_1$ and $T_2$.}
\label{fig:GNN_accuracy}
\end{figure}

\begin{figure*}[t]
\centering
\includegraphics[width=\textwidth]{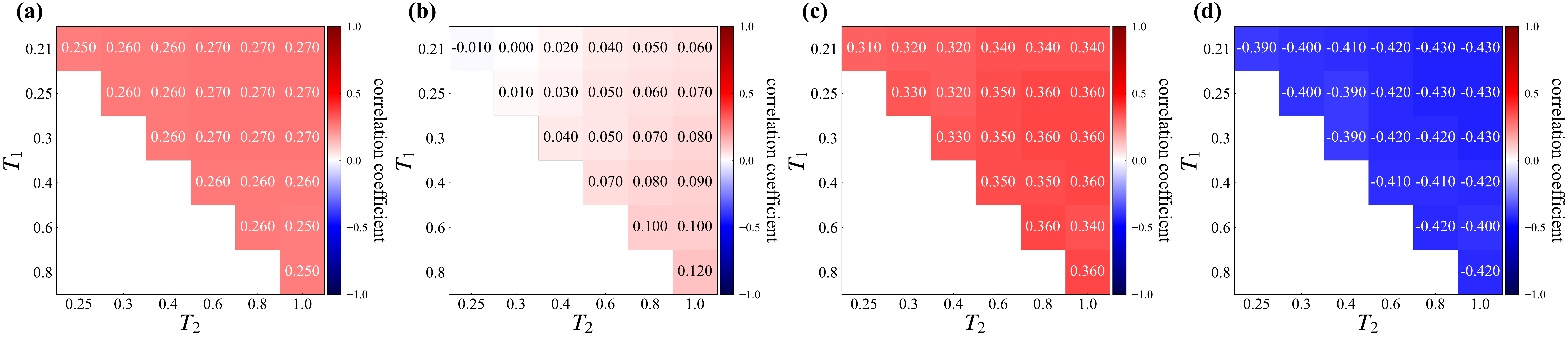}
\caption{Pearson correlation coefficients between node features,
 denoted as $H'_u$, generaged by GNN, and structural descriptors
 including two BOOs, $Q_6$ (a), $Q_4$ (b), Voronoi cell
 volume (c), and C.N. (d) for each particle.
The temperature combinations are identical to those presented in Fig.~\ref{fig:GNN_accuracy}.
}
\label{fig:GNN_node_feagure}
\end{figure*}

\begin{figure*}[t]
\centering
\includegraphics[width=\textwidth]{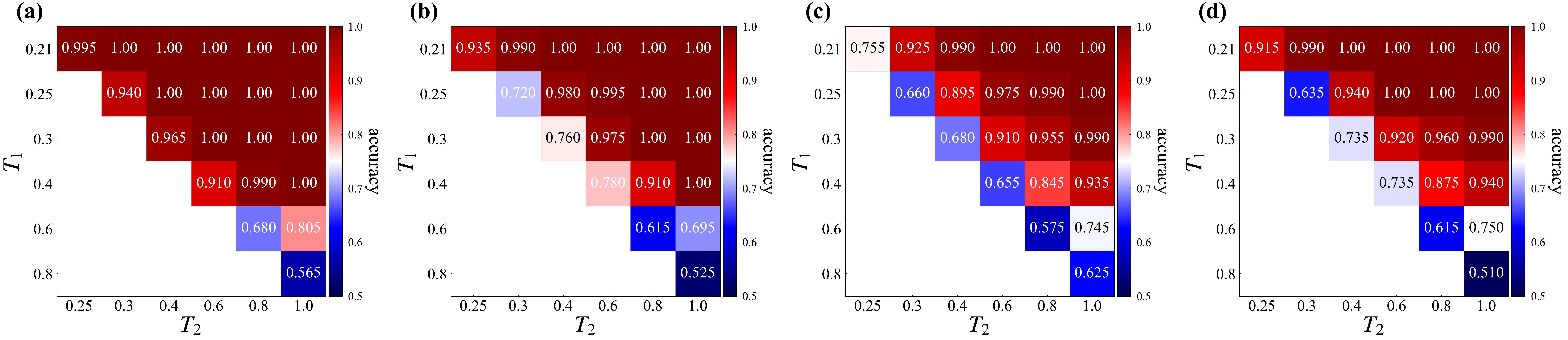}
\caption{Fraction of FCNN
 prediction accuracy for 
classification between two temperatures, $T_1$ and $T_2$, using two
 BOOs, Q6 (a), Q4 (b), Voronoi cell volume (c), and 
 C.N. (d) for each particle.
}
\label{fig:FCNN_accuracy}
\end{figure*}

\begin{figure*}[t]
\centering
\includegraphics[width=\textwidth]{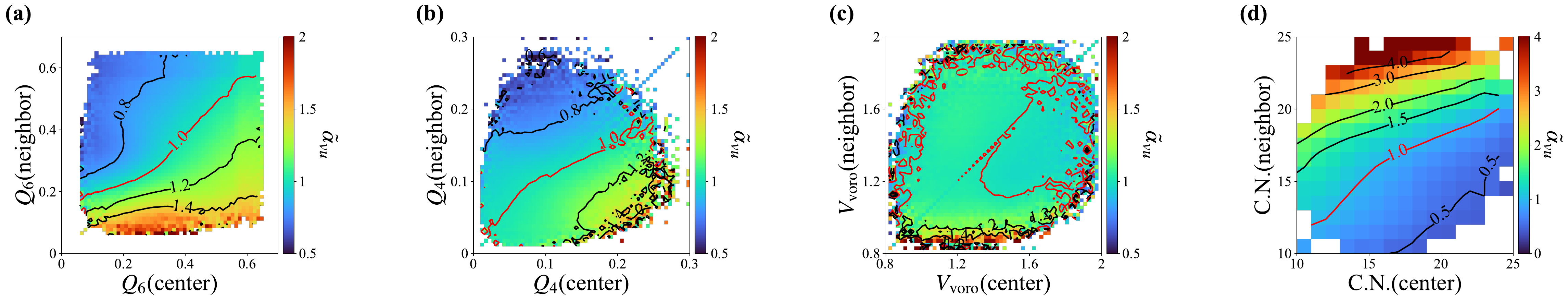}
\caption{Probability distribution plots using two BOOs, $Q_6$ (a) and $Q_4$ (b), 
Voronoi cell volume (c), and C.N. (d) for temperature combination of $T_1=0.21$ and $T_2=1.0$.
The distributions are colored by the corresponding coordination-weighted
 attention coefficient
 $\tilde{\alpha}_{vu}$ values given in the color bar.
}
\label{fig:attention_map}
\end{figure*}

\section{Results and discussion}

\subsection{GNN classification of different temperatures}
\label{result:GNN}

Figure~\ref{fig:GNN_accuracy}(a) and (b) present the prediction accuracy 
and the receiver operating characteristic area under the curve (ROC-AUC,
hereafter referred to simply as AUC)
by our structure-based GNN temperature predictions, evaluated 
across various combinations of two temperatures, $T_1$ and $T_2$, respectively.
In each combination of $T_1$ and $T_2$, $T_1$ is selected to be a lower
temperature than $T_2$.
Here, in the interpretation of prediction output, 
we assign the label $T_1$ if the output yields a value less than 0.5,
and $T_2$ otherwise.
The results demonstrate that the structure at each of the two
temperature combinations
examined in this study can be
perfectly distinguished from one another.
The evolution of the epoch-dependent loss function for both the training
and test datasets is depicted in Fig.~S3 of the supplementary material.
Notably, the GNN successfully classifies
different temperatures, even when
 the radial distribution function $g(r)$ shows small
 variations, while the $\alpha$-relaxation time $\tau_\alpha$ 
 increases significantly 
 as the temperature decreases
(see Fig.~S1(a), (b), and (c) of the supplementary material).
However,
a slight decrease in the prediction accuracy was observed
for combinations of higher temperatures, \textit{e.g.}, $T_1=0.8$
and $T_2=1.0$.
This result implies that 
the GNN predictions are accompanied by a specific level of resolution in the structure
classification across varying temperature.
To explain this aspect further, we analyzed the prediction 
accuracy and AUC of GNNs for supercooled
states within the temperature range 0.21-0.25, as shown in 
Fig.~S4 of the supplementary material, with the 
corresponding epoch-dependent loss function presented in Fig.~S5 of the supplementary material.
A decline in 
predictive performance was particularly evident for
closely spaced temperature combinations, consistent with insufficient
optimization of loss function.
These findings suggest that closely spaced temperatures result in
overlapping fluctuations in the potential energy, making it challenge
for GNNs to distinguish between structures between two different
temperatures (see Fig.~S1(d) of the supplementary material).
This limitation may be improved by using inherent structures obtained through
energy minimization, which would reduce reducing energy overlap.


\subsection{Node features generated by GNN}

We next examined the physical properties of the node features generated
by the GNN, denoted as $H'_u$
for the node index $u$ in Fig.~\ref{fig:GNN}, which were
were subsequently employed in the FCNN to predict temperature.
For this aim, we calculated BOOs ($Q_6$ and $Q_4$), 
Voronoi cell colume, and C.N. for each particle of a specific
configuration and compared these
structural descriptors with
the GNN-generated node features $H'_u$ using the same configuration.
Specifically, the Pearson correlation coefficient was calculated for
each 
$(T_1, T_2)$ pair snapshot and averaged over 200 configurations
(100 configurations from $T_1$ and 100 configurations of $T_2$
from the test dataset).
The results are
illustrated in Fig.~\ref{fig:GNN_node_feagure}.
Overall, $Q_6$ and Voronoi cell volumes display positive
correlations with the node
features reaching values approximately 0.3, while C.N. exhibit negative
correlations of up to 0.4 in magnitude.
Note that $Q_4$ also shows slight positive correlations, though less procounced
than those of $Q_6$.
A more detailed observation reveals that, 
for all ($T_1$, $T_2$) pairs, the absolute values of 
the correlations with node features
generated by GNNs follow the order: C.N. $>$ Voronoi cell volume $>$ $Q_6$ $>$ $Q_4$.
This trend can be attributed to the methodology used to compute structural
descriptors:
$Q_6$ and $Q_4$ incorporate angular information of neighboring
particles.
In contrast, C.N. and Voronoi cell volume provide a more direct capture
of local structural environment.
The observed correlations between node features and structural descriptors suggest
that 
the GNN's success in two-temperature discrimination arises from its
ability to effectively capture 
insights into the local particles environment.

\subsection{FCNN classification of different temperatures using
structural descriptors}
\label{result:FCNN}

The correlations between GNN-generated node features and
physically-defined structural descriptors suggest that 
ML with the FCNN, utilizing these structural descriptors, may
also effectively 
discriminate between the two temperatures.
In fact, the distribution of each structural
descriptor exhibits some temperature dependence, as shown in Fig.~S2 of
the supplementary material, suggesting that FCNN
can presumably distinguish temperature differences.

We thus implemented an FCNN to classify the structure between two temperatures.
The architecture consists of an input layer with 4096 nodes, a single
hidden layer also with 4096 nodes utilizing the ReLU,
and a binary output layer predicting the temperature ($T_1$ or
$T_2$) via a sigmoid function. 
The model is trained by minimizing binary cross-entropy. 
Input variables for each particle include two BOOs ($Q_6$ and $Q_4$), Voronoi
cell volume, and C.N. 
This FCNN architecture mirrors the final component of the GNN (see Fig.~\ref{fig:GNN}).
Similar to the GNN, we employed PyTorch with the Adam optimization
algorithm, employing a learning rate of $10^{-6}$ and a batch size of 128.
The evolution of the epoch-dependent loss function for both the training
and test datasets
is depicted in Supplementary Figs.~S6-S9.
Similar to Fig.~\ref{fig:GNN_accuracy}(a), Fig.~\ref{fig:FCNN_accuracy}
depicts the fraction of accurate predictions by our FCNN
for classification between
two temperatures, $T_1$ and $T_2$, using two BOOs, $Q_6$ (a) and $Q_4$
(b), Voronoi cell volume (c), and C.N. (d).
The results of AUC are shown in Fig.~S10(a)-(d) of the supplementary material.

When $Q_6$ is used as the input variable, the structure
at the lowest temperature, $T_1 = 0.21$, is clearly distinguished
from those at other temperatures.
However, distinguishing 
structures between high temperatures proves challenging, in contrast to the 
GNN results.
When other structural descriptors, $Q_4$, Voronoi cell
volume, and C.N., are used as input variables, the predictive performance
follows a similar trend to that of $Q_6$, albeit with 
an overall degradation.
These reductions in performance, particularly in the high-temperature
regime, 
suggest that using physically-defined structural descriptors as input
variables in the FCNN presents certain limitations.
In particular, the results suggest that FCNNs with
structural descriptors are more sensitive to thermal
fluctuations at high temperatures. 
In contrast, the decrease in prediction accuracy for GNNs is relatively
modest, implying that the GNN effectively suppresses the influence of
thermal fluctuations and generates more robust features through the
message passing process.



\subsection{Attention coefficient and comparison with structural descriptors}
\label{result:attention}

The self-attention layer of our GNN architecture quantifies the
normalized attention coefficient $\alpha_{vu}$ (see Fig.~\ref{fig:GNN}).
As outlined in the Introduction, this coefficient $\alpha_{vu}$ evaluates the
importance of node $u$'s feature to 
node $v$ 
in the graph structure.
Specifically, strong relationships between nodes are
reflected in larger attention coefficients, offering insights into 
the inter-particle correlations within amorphous structures.
Here, 
the normalized attention coefficient, $\alpha_{vu}$, multiplied by the number of neighboring nodes of $v$,
including node $v$ itself, is
analyzed 
as the coordination-weighted attention
coefficient, $\tilde{\alpha}_{vu}$
(see Eq.~\eqref{eq:attention} in ``Mehtods'' for the definition of
$\alpha_{vu}$ using the Softmax function).
This multiplication 
accounts for the variations in the number of
neighboring nodes for each node.
Thus, a value of unity for $\tilde{\alpha}_{vu}$ represents the average
level of 
the normalized attention coefficient, $\alpha_{vu}$.

It is essential to note that the attention coefficient
solely serves as the foundation basis for 
the GNN predictions, even though it currently lacks explicit physical
interpretation.
Therefore, a comparative analysis between the
attention coefficient and structural descriptors including BOOs ($Q_6$
and $Q_4$), Voronoi cell volume, and C.N., is significant.
This analysis aims to provide deeper physical understanding into the high
predictive performance of the GNN.
To achieve this, 
we examine the correlations between structural descriptors of a given
particle (node $v$)
of interest and its edge-connected neighboring particle (node $u$), characterized by the
corresponding 
coordination-weighted attention coefficient, $\tilde{\alpha}_{vu}$.

Figure~\ref{fig:attention_map} presents
two-dimensional probability distributions for two BOOs, $Q_6$ (a), $Q_4$ (b), Voronoi cell volume,
and C.N. (d) for the two temperatures, $T_1=0.21$ and $T_2=1.0$. 
These are averaged over 200 configurations
(100 configurations from $T_1$ and 100 configurations of $T_2$
from the test dataset).
The horizontal axis represents values for 
a given central particle (node $v$), while the vertical axis corresponds
to values for 
its neighboring particle (node $u$).
The distributions are further colored based on the
corresponding $\tilde{\alpha}_{vu}$, averaged over the number of particle pairs stored in each
two-dimensional bin.
$\tilde{\alpha}_{vu}=1$ is also highlighted in red.

Note that 
the shape of the distribution is also
symmetric, reflecting the inherent symmetry of 
structural descriptors between two particles.
However, 
due to the asymmetry, $\tilde{\alpha}_{vu}$ and $\tilde{\alpha}_{uv}$, 
with respect to two nodes $v$ and $u$, as described in
``Methods'', the color map exhibits asymmetry when considering both
$\tilde{\alpha}_{vu}$ and $\tilde{\alpha}_{uv}$ by exchanging the roles of the tow particles.
The results of other temperature combinations are shown in
Figs.~S11-S14 of the supplementary material.

Remarkably, Fig.~\ref{fig:attention_map}(a) illustrates that 
neighboring particles with small
$Q_6$
exhibit larger values ($\tilde{\alpha}_{vu}>1$).
This finding suggests that 
neighboring particles with more disordered orientation
play a significant role in the local environment of the central particle.
A similar trend is observed for $Q_4$, though the contrast is less
pronounced compared to $Q_6$, as shown in Fig.~\ref{fig:attention_map}(b).
In contrast, 
correlations with Voronoi cell volume are weaker (see Fig.~\ref{fig:attention_map}(c)).
Furthermore, Fig.~\ref{fig:attention_map}(d) reveals that 
neighboring particles with 
large C.N. values also exhibit 
larger values ($\tilde{\alpha}_{vu}>1$).
This trend suggests that, similar to $Q_6$, 
central particles tend to focus on disordered
orientations in their neighboring environment.

These results demonstrates that the GNN effectively identifies static
heterogeneities between locally disordered
structures, enabling it to distinguish between the two temperatures.
In particular, the correlation between coordination-weighted attention coefficients and
structural descriptors, particularly $Q_6$ and C.N., 
arises from how the
relative coordinates of neighboring particles for a given central particle are encoded into
the graph data utilized by the GNN.

\section{Conclusions}

In this study, we utilized ML, specifically
GNNs, to tackle the problem of classifying structural
configurations generated by MD simulations of three-dimensional glass-forming liquids at various
temperatures. 
In contrast to previous studies,~\cite{swanson2020Deep, oyama2023What,
liu2024Classification} which utilize ML to differentiate
between vitrified and liquid structures at a constant cooling rate, our
study developed a GNN model capable of distinguishing between two
different temperatures within a branch of the supercooled state.

The GNNs have yielded insights into the
classification of structural configurations in glass-forming liquids by
varying temperatures. 
Notably, we found that the structure of different temperatures with
significantly different $\alpha$-relaxation times 
can be perfectly distinguished, while
the ability to discriminate
between states with close potential energies was degraded. 
This result suggests a convincing relationship between structural
characteristics and slow dynamics observed in glass-forming liquids.

The adoption of GNNs proved highly advantageous, given
their inherent capability to process graph data efficiently. 
By representing individual particles as nodes and their connections as
edges within our graph-based framework, we utilized the advantages of
GNNs to automatically generate feature
representations, without relying on prior physical information.
These features
captured intricate relationships with neighboring particles and allowed
us to discern distinct temperature-dependent structures.
Notably, node features generated by the GNN exhibit correlations
with physically-defined structural descriptors including the BOOs, $Q_6$
and $Q_4$, Voronoi cell volume, and C.N.
The results demonstrates that the GNN learns information related to the
local structure of the particle configuration from the amorphous
structure.

Furthermore, we conducted a comparative analysis by applying
conventional FCNNs to the same binary classification problem. 
Using $Q_6$, $Q_4$, Voronoi cell volume, and C.N. as input variables,
this approach aimed to provide deeper insights into the fundamental
mechanisms underpinning ML in glass-forming liquids.
This analysis not only corroborated the significance of structural
descriptors but also
highlighted the distinct strengths of GNNs in capturing 
important features from amorphous configurations.

Finally, our investigation unveiled the significance of incorporating
the self-attention mechanism into the GNN architecture, shedding light on
the inter-node relationships within the system. 
This self-attention mechanism, originally introduced in Graph Attention Networks,
enabled us to assess the importance of each node's connections with its
neighbors, quantifying the features associated with each node. 
This human-interpretable model was crucial in understanding the
structural changes occurring with temperature variations. 
Swanson \textit{et al.} reported
that, in two-dimensional systems, 
edges with large attention coefficients 
tend to 
form a large connected graph in the glass, while several smaller
disconnected graphs are observed in the liquid state~\cite{swanson2020Deep}.
By contrast, we provide physical interpretations to the
coordination-weighted attention coefficient, $\tilde{\alpha}_{vu}$.
Notably, our results reveal that small $Q_6$ values and large C.N. values
of neighboring particles emerge as key features, both indicative of
locally disordered structures.
These findings emphasize the need for 
further analysis to deepen our understanding of the compelling
relationship between static structural features and the dynamic
behaviors of individual particles, aligning with the direction of studies by
Bapst \textit{et al.} and others~\cite{jung2025Roadmap}.

Our exploration of GNNs for structural
classification in glass-forming liquids opens several promising aspects
for future research. 
Extending this work to encompass a broader range of
materials, including diﬀerent glass formers such as the widely used
Kob–Andersen model, silica glasses, polymer glasses and supercooled
water, will contribute to a more comprehensive understanding of the
relationships between structural features and dynamics. 
In particular,
the integration of additional structural descriptors
parameters~\cite{tong2018Revealinga, tong2019Structural}, in conjunction
with the GNN+self-attention framework, will valuable insights into the underlying mechanisms in glass-
forming liquids.
For silica glasses and supercooled water, combining this approach with 
structural descriptors, which captures the degree of local
tetrahedral ordering, may provide deeper insights into the structural
origins of their anomalous behaviors.~\cite{shi2018Microscopic}
In addition, incorporating persistent homology
could further advance the analysis from a
topological perspective, even in polymer glasses lacking a well-defined
structural motif.~\cite{hiraoka2016Hierarchical}

\section*{SUPPLEMENTARY MATERIAL}

The supplementary material includes 
temperature dependence of $g(r)$, $F_\mathrm{s}(k, t)$, $\tau_\alpha$,
and potential energy (Fig.~S1); 
probability distribution of structural descriptors, including the
 BOOs, Voronoi cell volume, and
 C.N. (Fig.~S2); 
epoch-dependent loss function of GNNs (Fig.~S3); 
fraction of GNN accuracy and AUC for the temperature range 0.21-0.25 (Fig.~S4); 
epoch-dependent loss function of GNNs for the temperature range 0.21-0.25 (Fig.~S5);
epoch-dependent loss function of FNCCs using structural descriptors
(Fig.~S6, S7, S8, and S9);
fraction of FCNN AUC (Fig.~S10);
Probability distribution plots using structural descriptors for two
temperature combination, shown alongsidethe coordination-weighted
attention coefficient, $\tilde{\alpha}_{vu}$, for comparison (Fig.~S11, S12, S13, and S14).

\begin{acknowledgments}
We express our gratitude to H.~Shiba for generously providing the
 information regarding the PyG\_BOTAN code~\cite{PyG_BOTAN}.
We also extend our thanks to N.~Oyama for opening the code used in Ref.~\onlinecite{oyama2023What}.
We acknowledge T. Kawasaki and H. Mizuno for their insightful discussions.
This work was supported by 
JSPS KAKENHI Grant-in-Aid 
Grant Nos.~\mbox{JP25K00968}, \mbox{JP24H01719}, \mbox{JP22H04542}, \mbox{JP22K03550}, and \mbox{JP23H02622}.
We acknowledge support from
the Fugaku Supercomputing Project (Nos.~\mbox{JPMXP1020230325} and \mbox{JPMXP1020230327}) and 
the Data-Driven Material Research Project (No.~\mbox{JPMXP1122714694})
from the
Ministry of Education, Culture, Sports, Science, and Technology.
The numerical calculations were performed at Research Center of
Computational Science, Okazaki Research Facilities, National Institutes
of Natural Sciences (Projects: 24-IMS-C051 and 25-IMS-C052) and at the D3 Center,
 The University of Osaka.
\end{acknowledgments}

\section*{AUTHOR DECLARATIONS}

\section*{Conflict of Interest}
The authors have no conflicts to disclose.

\section*{Data availability statement}

The data that support the findings of this study are openly available in
Zenodo at https://doi.org/10.5281/zenodo.15151938.
Further data that support the findings of this study are available from
the corresponding author upon request.

%

\end{document}


\setcounter{equation}{0}
\setcounter{figure}{0}
\setcounter{table}{0}
\setcounter{page}{1}

\noindent{\bf\Large Supplementary Material}
\vspace{5mm}
\begin{center}
\textbf{\large  
Graph neural network-based structural classification of glass-forming
 liquids and its interpretation via self-attention mechanism
\\
}

\vspace{5mm}
 
{Kohei Yoshikawa, Kentaro Yano, Shota Goto, Kang Kim, and Nobuyuki Matubayasi}
\\

\vspace{5mm}

\noindent
\textit{
Division of Chemical Engineering, Graduate School of Engineering Science, The University of Osaka, Osaka 560-8531, Japan}

\end{center}

\begin{figure}[h]
\centering
\includegraphics[width=0.75\linewidth]{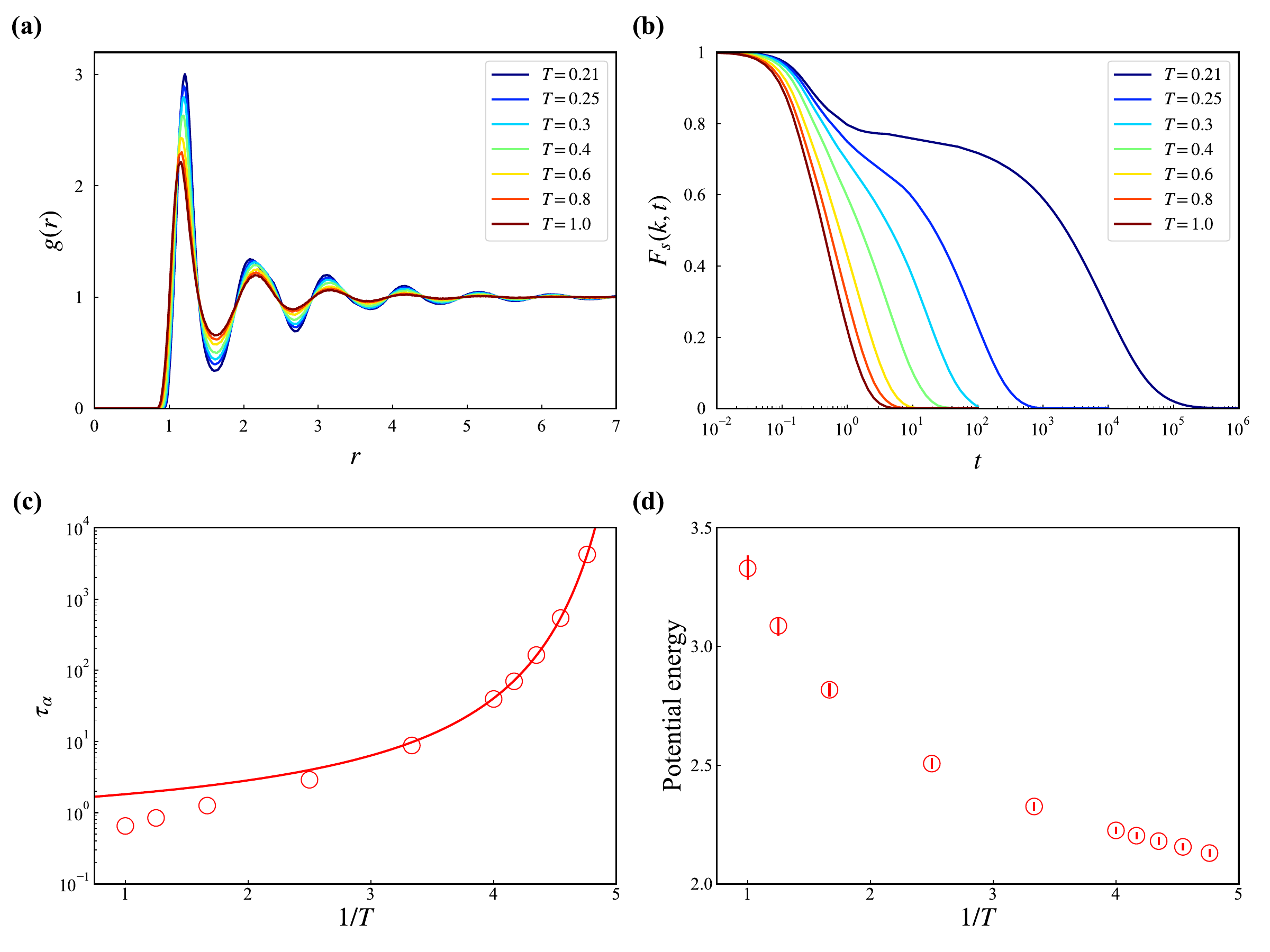}
\caption{(a) Radial distribution function $g(r)$ at
 various temperatures.
(b) Self-part of the intermediate scattering funciton $F_s(k, t)$ at
 various temperatures.
The wave number is chosen at $k=2\pi$.
In (a) and (b), particles species are not differentiated in the calculation of
 $g(r)$ and $F_s(k, t)$.
Note that the plots of $T=0.22-0.24$ are omitted for clarity. 
(c) Inverse temperature $1/T$ dependent $\alpha$-relaxation time
 $\tau_\alpha$ determined from the self-part of the
 intermediate scattering function,where $F_s(k, t=\tau_\alpha)=1/e$ at $k=2\pi$.
The red curve represents the Vogel--Fulcher--Tammann equation,
 $\tau_\alpha = \tau_0 \exp(D/ (T-T_\mathrm{K}))$, with three fitting
 parameter values of $\tau_0=1.34$, $D=0.235$, and $T_\mathrm{K}=
 0.181$.
(d) Inverse temperature $1/T$ dependent potential energy.
Error bars indicate the maximum and minimum values.
}
\end{figure}

\newpage
\begin{figure}[H]
\centering
\includegraphics[width=0.75\linewidth]{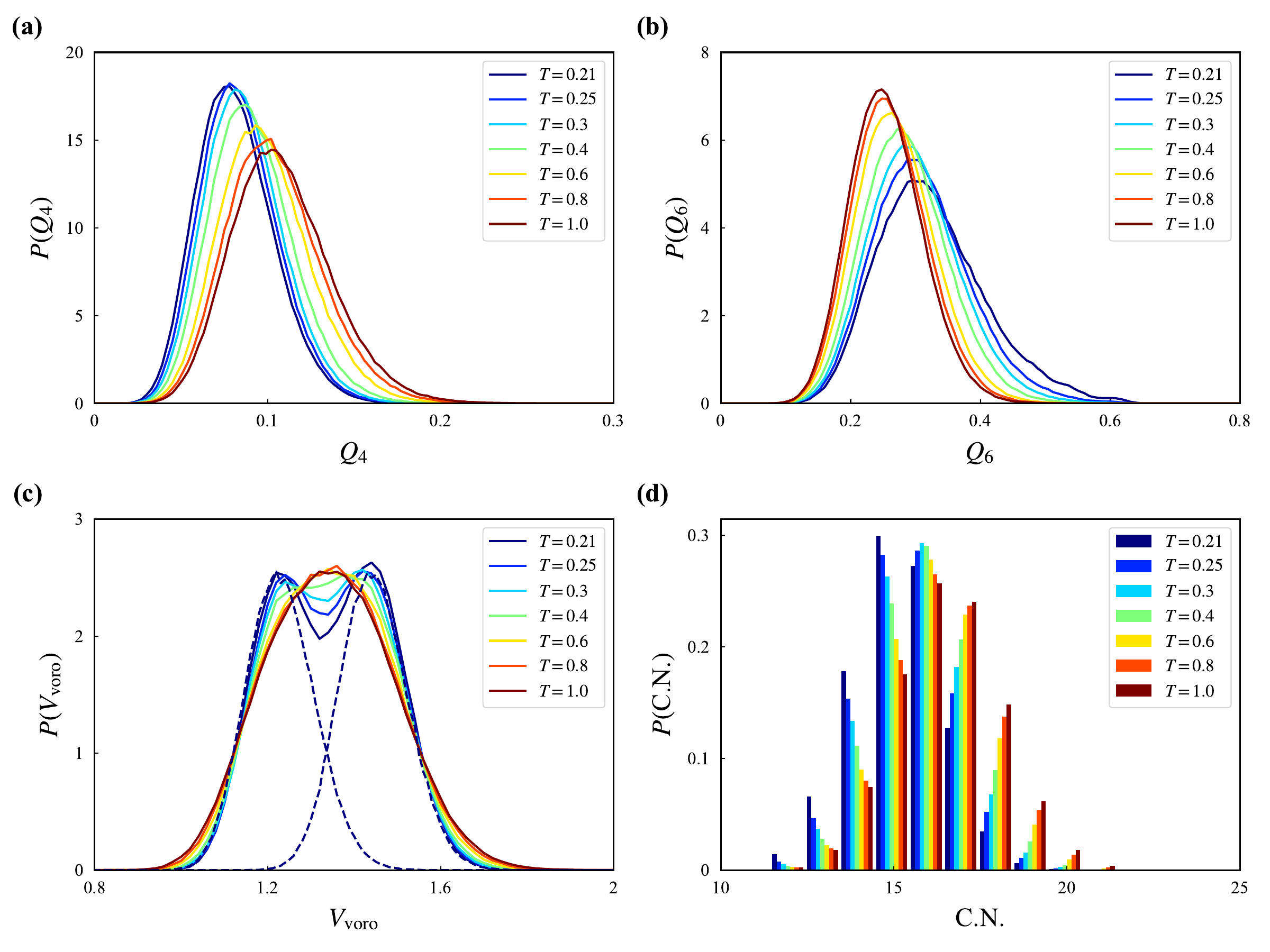}
\caption{Probability density distributions of $Q_6$ (a), $Q_4$ (b),
 Voronoi cell volume (c), and C.N. (d) at various temperatures.
In (c), at $T=0.21$, the distributions of small (type 1) and large (type
 2) particles are
 additionally plotted as dashed vurves.
Note that the plots of $T=0.22-0.24$ are omitted for clarity. 
}
\end{figure}

\newpage
\begin{figure}[H]
\centering
\includegraphics[width=0.75\linewidth]{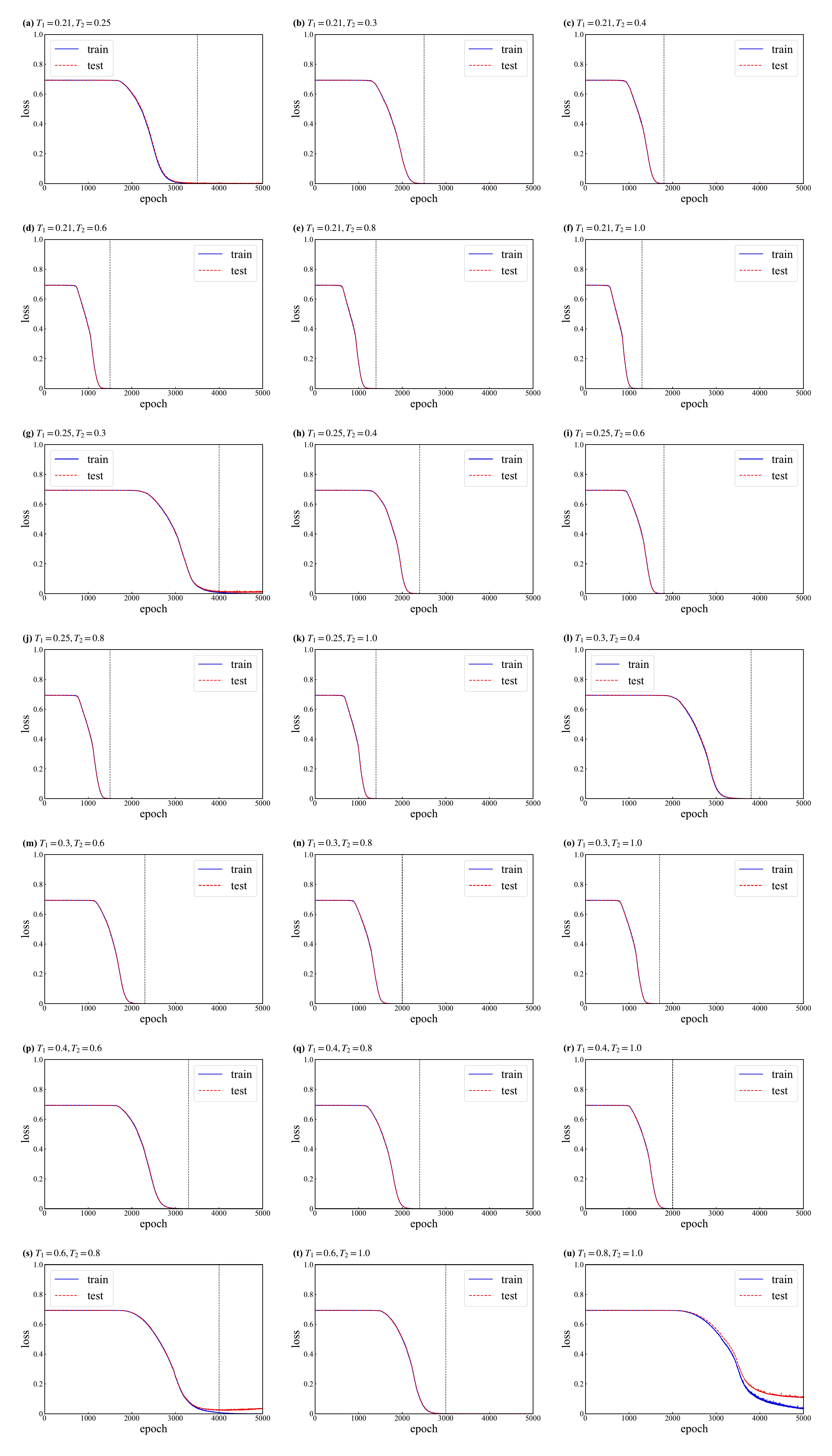}
\caption{
Epoch-dependent loss function of GNNs for training and test
 datasets for the structural classification between two temperatures,
 $T_1$ and $T_2$.
Vertical lines represent the number of epochs at which the learning was terminated.
If no vertical line is present, the training proceeded for the maximum number of epochs, up to 10000.
}
\end{figure}

\newpage
\begin{figure}[H]
\centering
\includegraphics[width=0.6\linewidth]{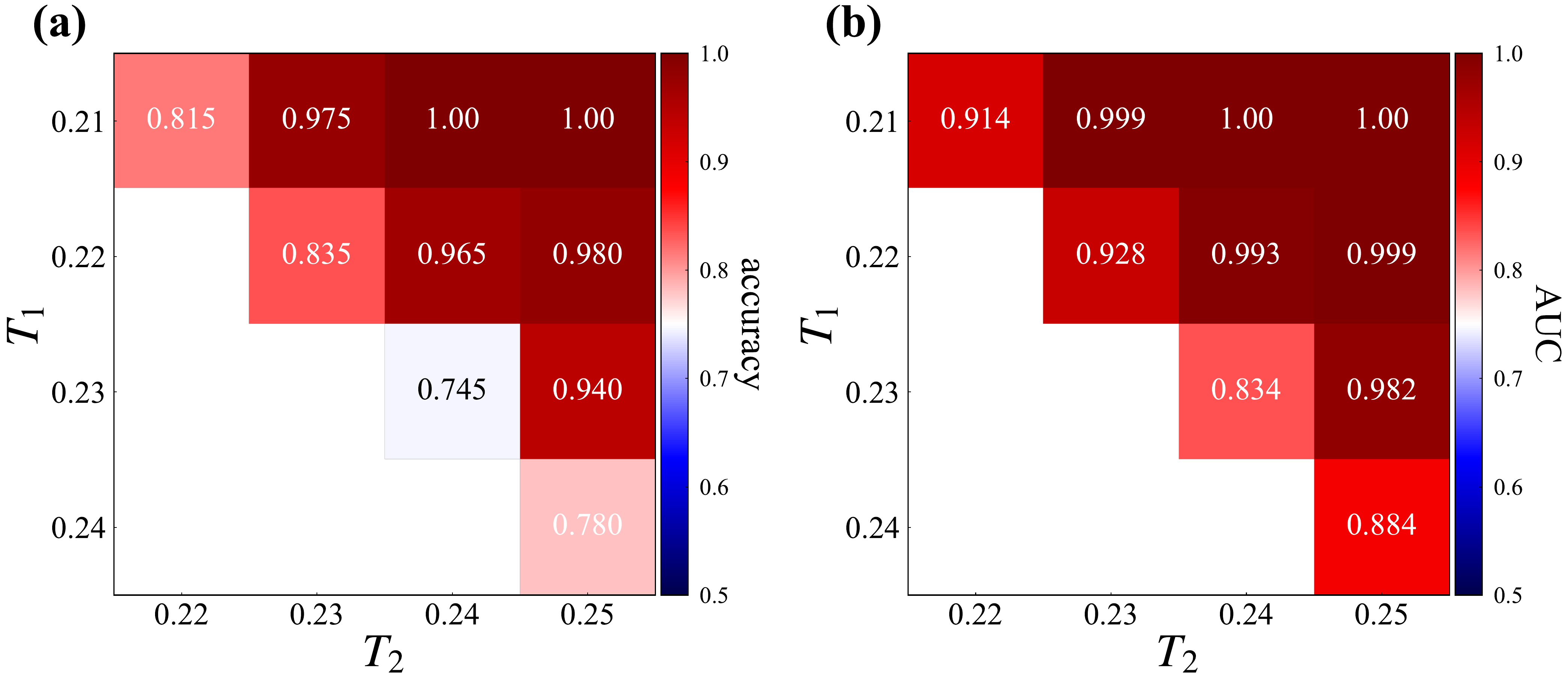}
\caption{Fraction of GNN prediction accuracy (a) and AUC (b) for
classification between two temperatures, $T_1$ and $T_2$, within the
temperature range $0.21-0.25$.}
\end{figure}

\begin{figure}[H]
\centering
\includegraphics[width=0.75\linewidth]{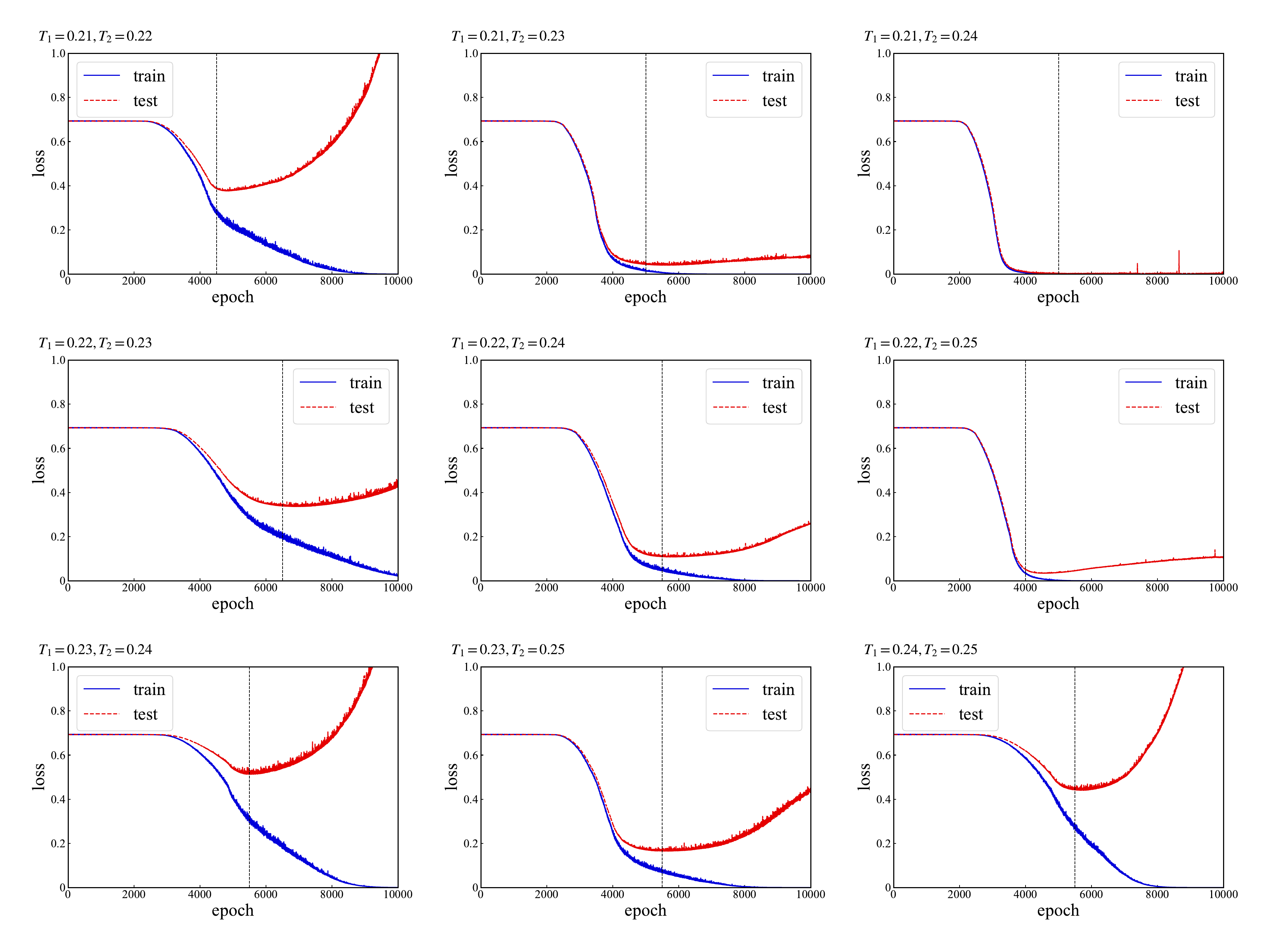}
\caption{
Epoch-dependent loss function of GNNs for training and test
 datasets for the structural classification between two temperatures,
 $T_1$ and $T_2$, within the
temperature range $0.21-0.25$.
Vertical lines represent the number of epochs at which the learning was terminated.
If no vertical line is present, the training proceeded for the maximum number of epochs, up to 10000.
}
\end{figure}

\newpage
\begin{figure}[H]
\centering
\includegraphics[width=0.75\linewidth]{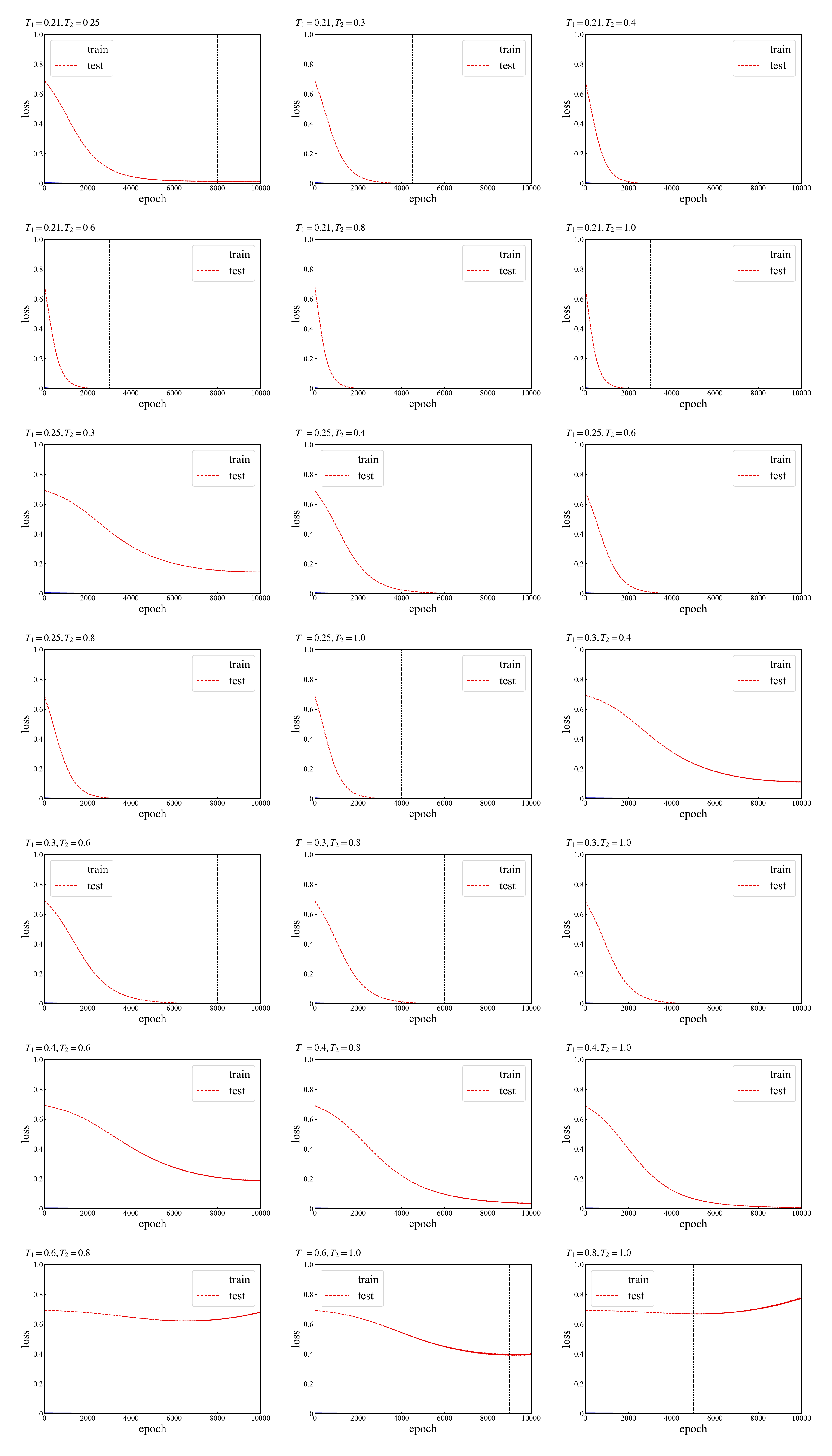}
\caption{
Epoch-dependent loss function of FNCCs using $Q_6$ for training and test
 datasets for the structural classification between two temperatures,
 $T_1$ and $T_2$.
Vertical lines represent the number of epochs at which the learning was terminated.
If no vertical line is present, the training proceeded for the maximum number of epochs, up to 10000.
}
\end{figure}

\newpage
\begin{figure}[H]
\centering
\includegraphics[width=0.75\linewidth]{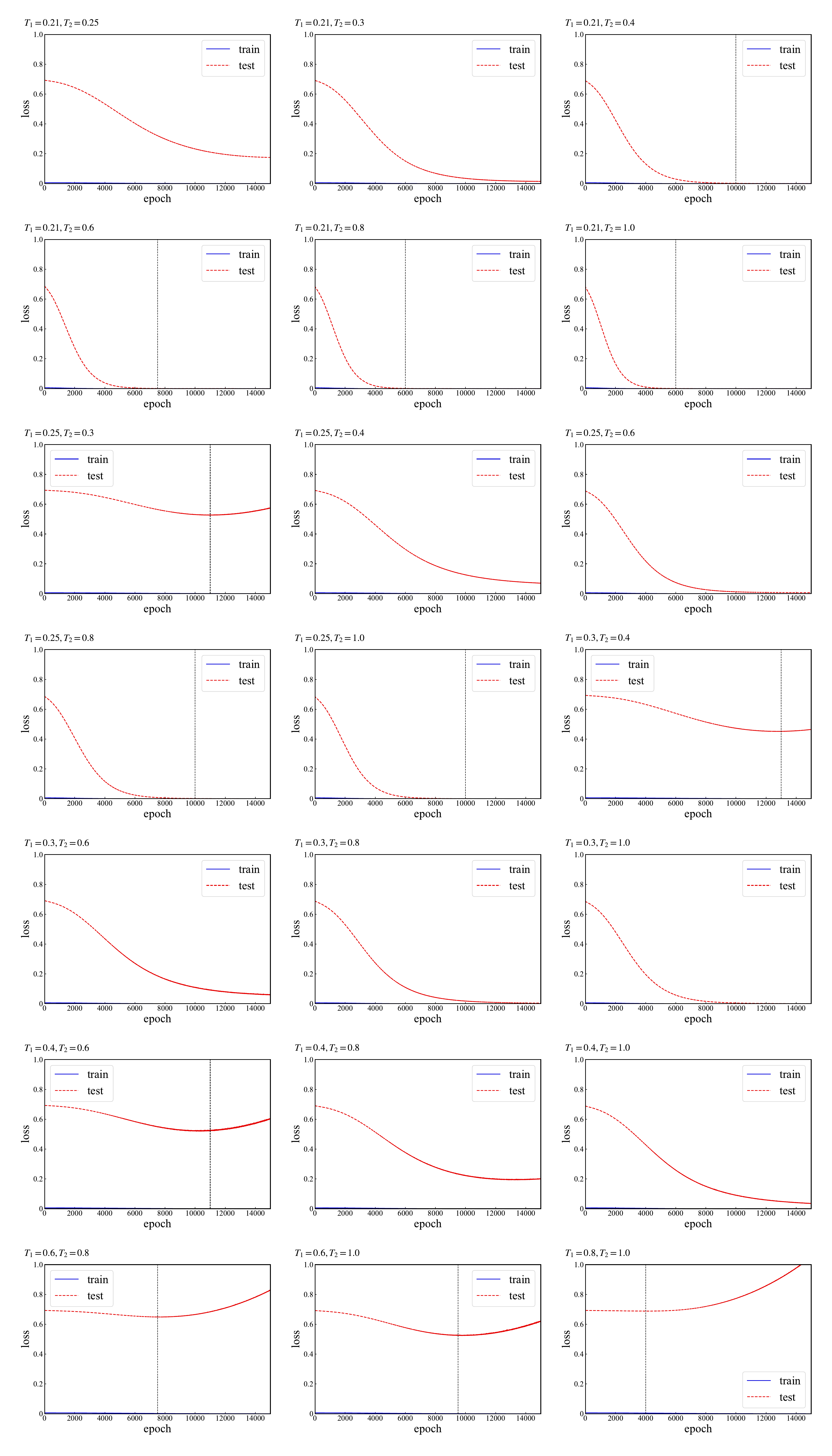}
\caption{
Epoch-dependent loss function of FNCCs using $Q_4$ for training and test
 datasets for the structural classification between two temperatures,
 $T_1$ and $T_2$.
Vertical lines represent the number of epochs at which the learning was terminated.
If no vertical line is present, the training proceeded for the maximum
 number of epochs, up to 10000.
}
\end{figure}

\newpage
\begin{figure}[H]
\centering
\includegraphics[width=0.75\linewidth]{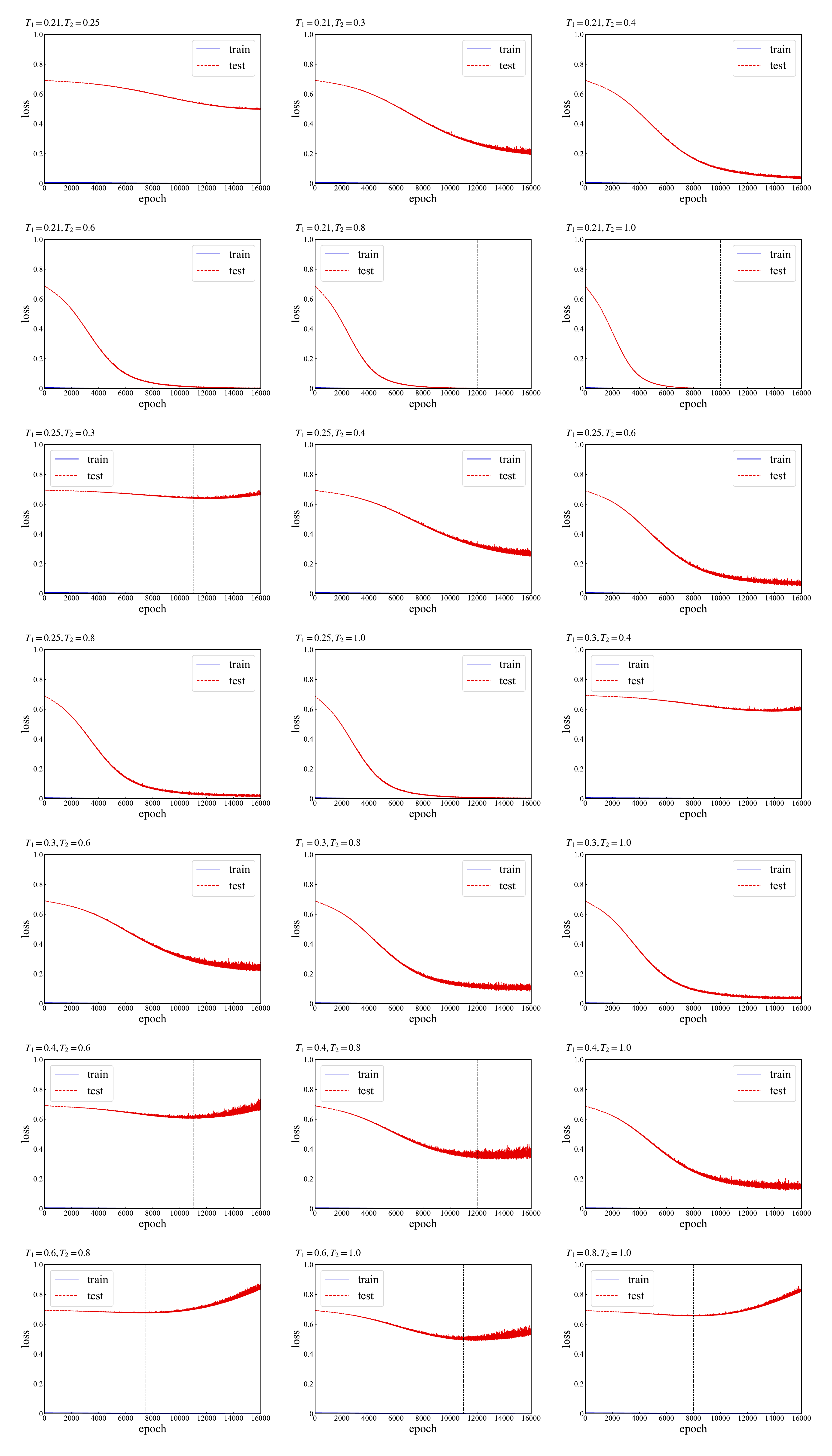}
\caption{
Epoch-dependent loss function of FNCCs using Voronoi cell volume for training and test
 datasets for the structural classification between two temperatures,
 $T_1$ and $T_2$.
Vertical lines represent the number of epochs at which the learning was terminated.
If no vertical line is present, the training proceeded for the maximum number of epochs, up to 10000.
}
\end{figure}

\newpage
\begin{figure}[H]
\centering
\includegraphics[width=0.75\linewidth]{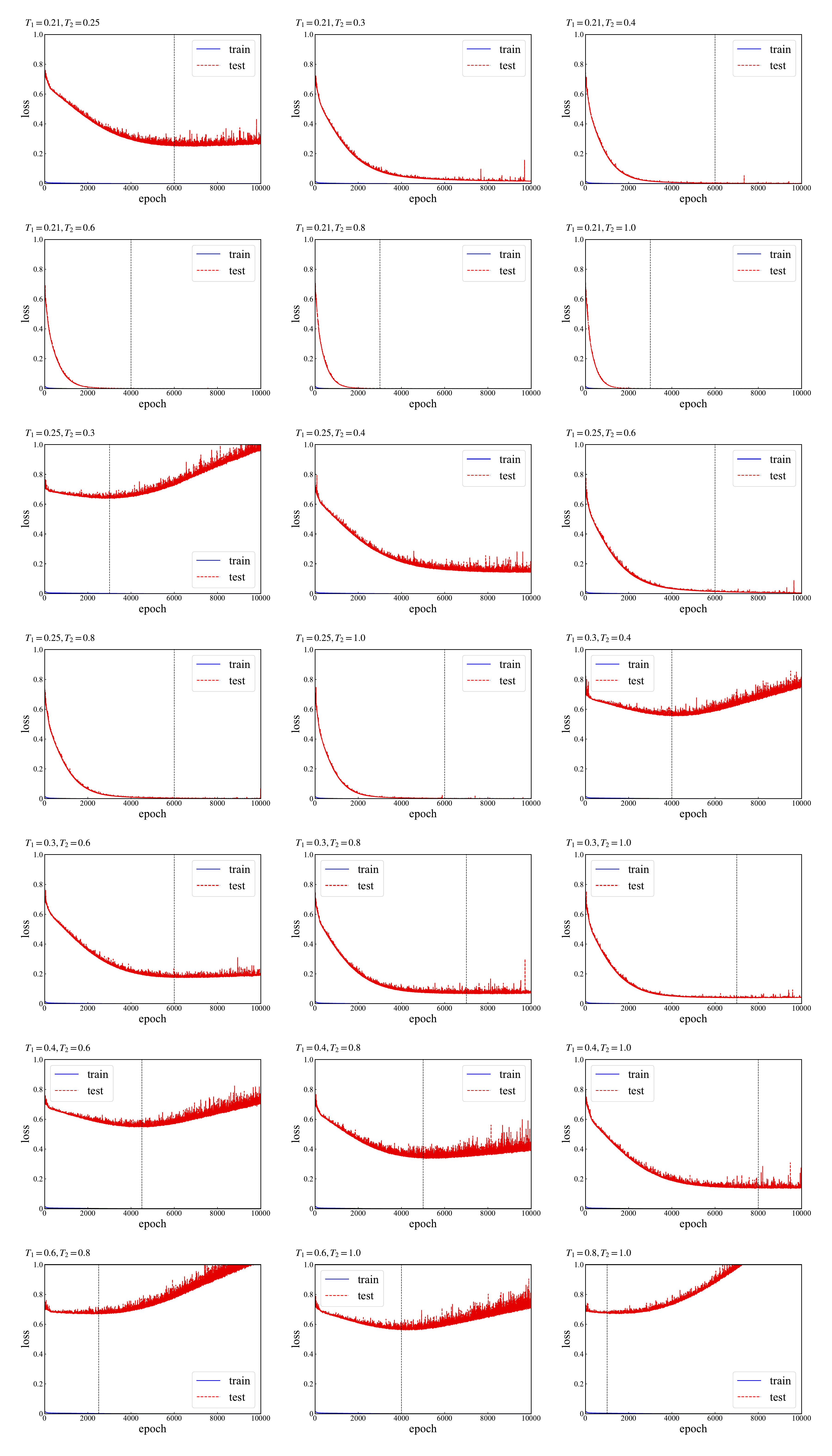}
\caption{Epoch-dependent loss function of FNCCs using C.N. for training and test
 datasets for the structural classification between two temperatures,
 $T_1$ and $T_2$.
Vertical lines represent the number of epochs at which the learning was terminated.
If no vertical line is present, the training proceeded for the maximum number of epochs, up to 10000.
}
\end{figure}

\newpage
\begin{figure}[H]
\centering
\includegraphics[width=0.9\linewidth]{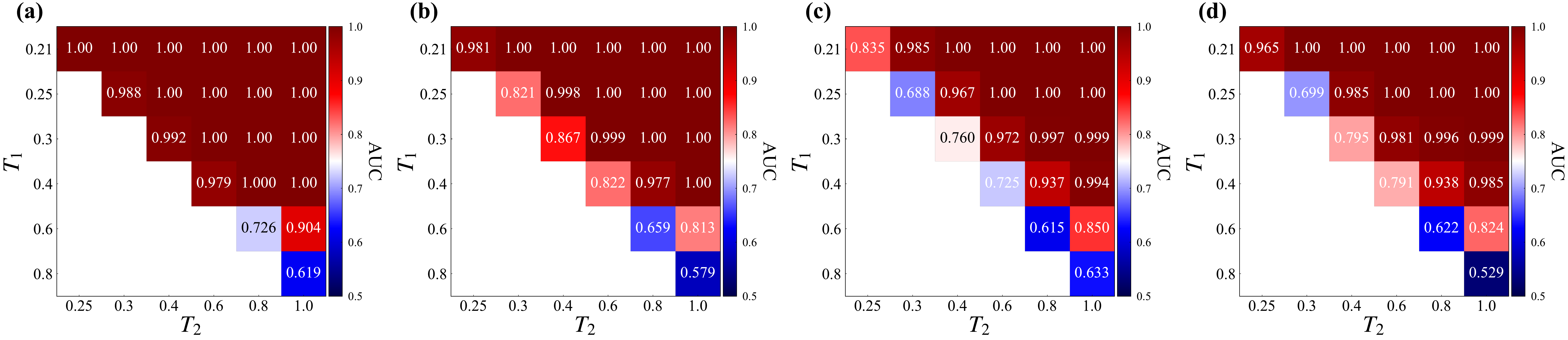}
\caption{Fraction of FCNN AUC for
classification between two temperatures, $T_1$ and $T_2$, using $Q_6$
 (a), $Q_4$ (b), Voronoi cell volume (c), and C.N. (d).}
\end{figure}

\begin{figure}[h]
\centering
\includegraphics[width=0.9\linewidth]{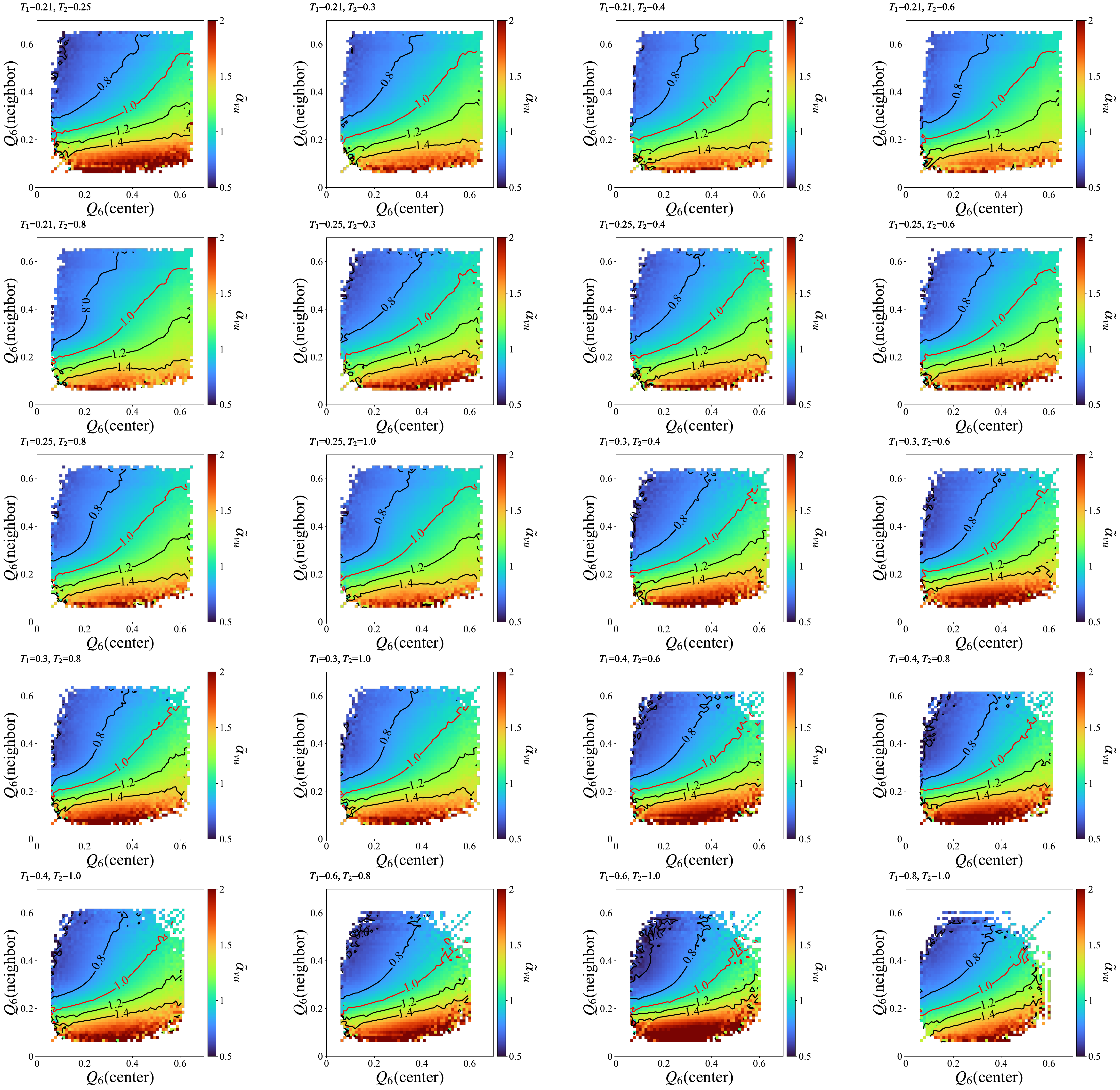}
\caption{Probability distribution plots using $Q_6$
for temperature combination of $T_1$ and $T_2$.
The distributions are colored by the corresponding coordination-weighted
 attention coefficient
 $\tilde{\alpha}_{vu}$ values given in the color bar.
}
\end{figure}

\newpage
\begin{figure}[H]
\centering
\includegraphics[width=0.9\linewidth]{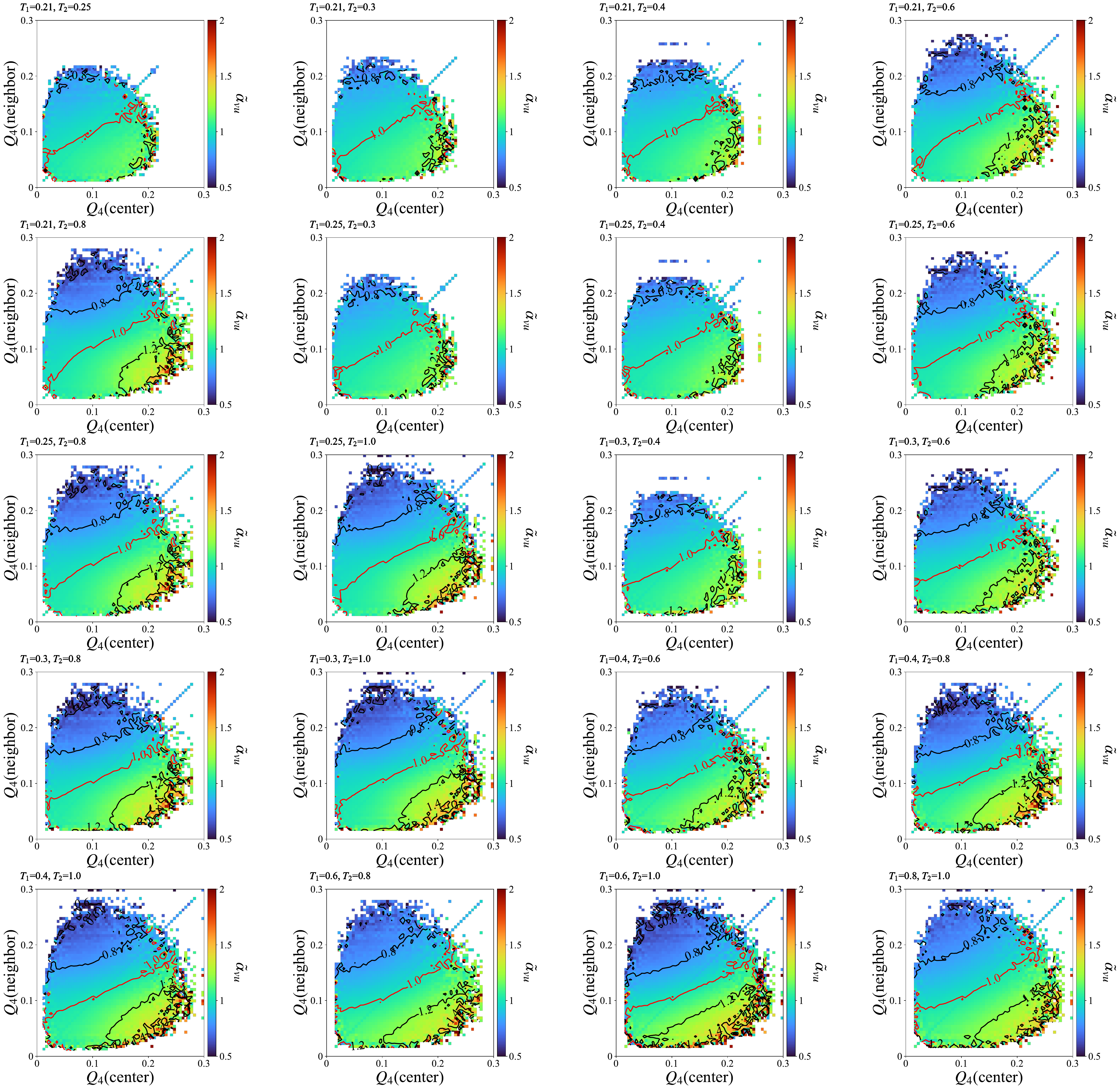}
\caption{Probability distribution plots using $Q_4$
for temperature combination of $T_1$ and $T_2$.
The distributions are colored by the corresponding coordination-weighted
 attention coefficient
 $\tilde{\alpha}_{vu}$ values given in the color bar.
}
\end{figure}

\newpage
\begin{figure}[H]
\centering
\includegraphics[width=0.9\linewidth]{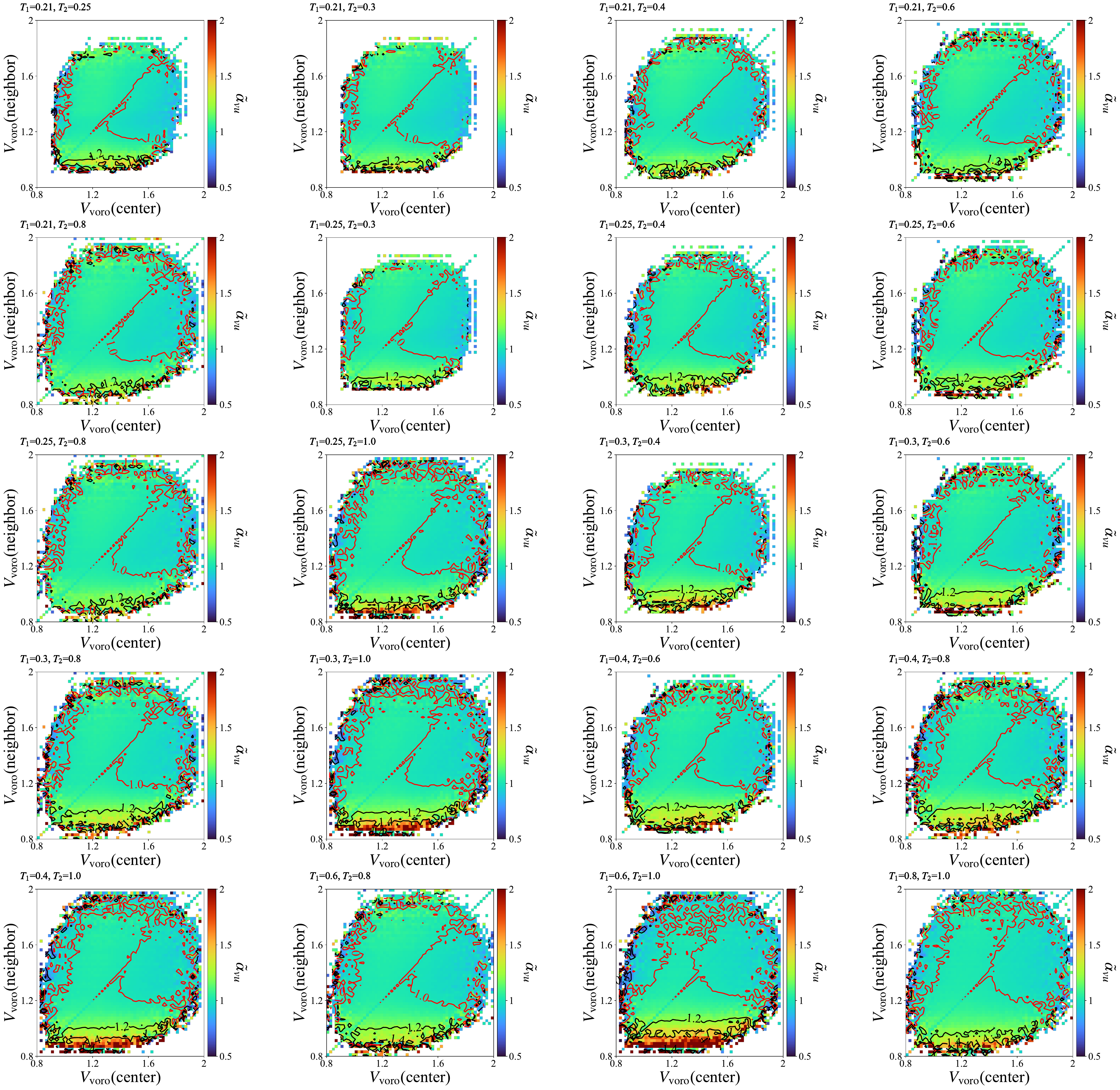}
\caption{Probability distribution plots using Voronoi cell volume
for temperature combination of $T_1$ and $T_2$.
The distributions are colored by the corresponding coordination-weighted
 attention coefficient
 $\tilde{\alpha}_{vu}$ values given in the color bar.
}
\end{figure}

\newpage
\begin{figure}[H]
\centering
\includegraphics[width=0.9\linewidth]{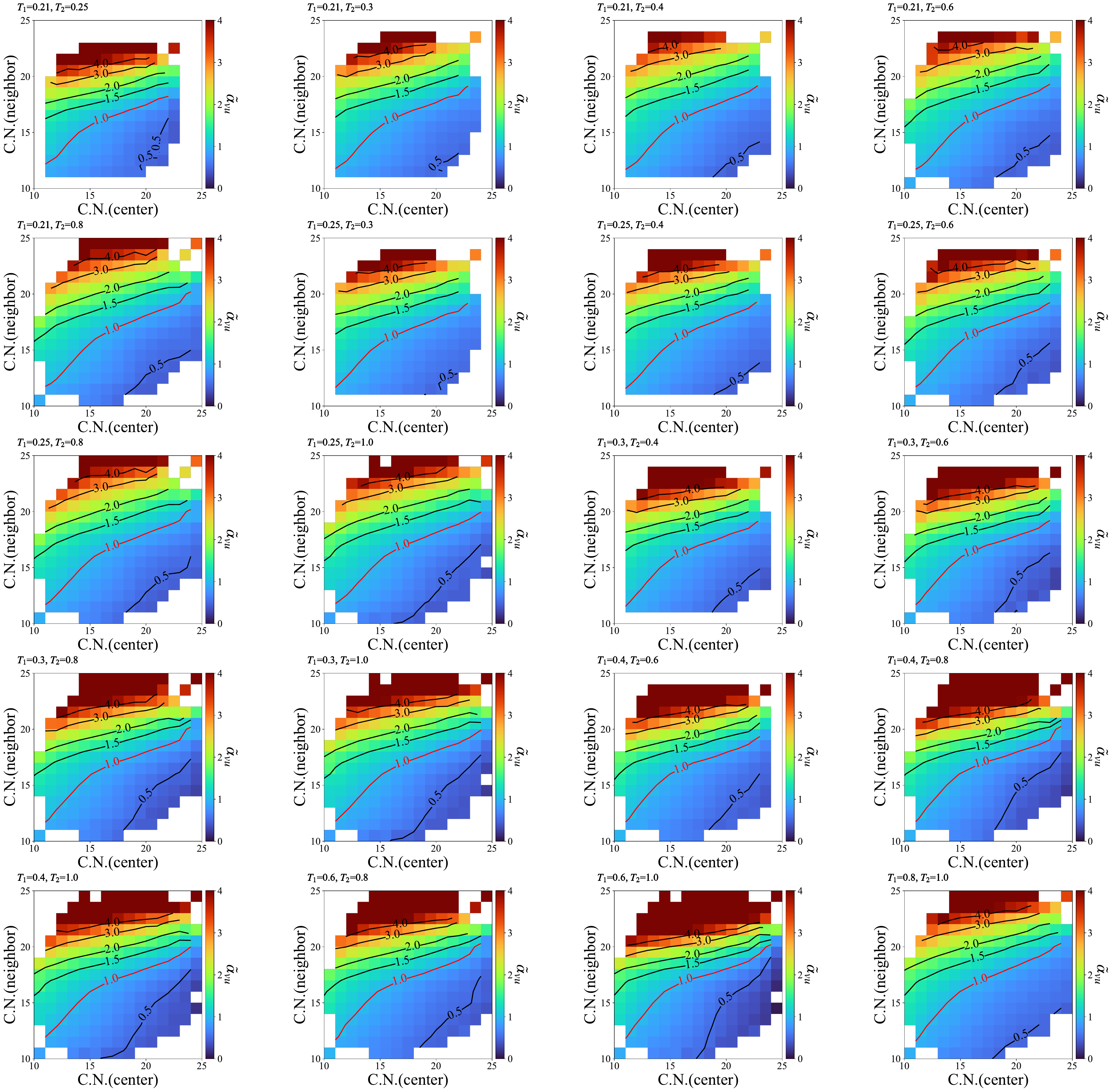}
\caption{Probability distribution plots using C.N.
for temperature combination of $T_1$ and $T_2$.
The distributions are colored by the corresponding coordination-weighted
 attention coefficient
 $\tilde{\alpha}_{vu}$ values given in the color bar.
}
\end{figure}